\documentclass[11pt]{article}
\usepackage{graphicx}
\usepackage{amssymb,amsmath}
\usepackage{epsfig}
\def\To{\rightarrow}

\def\C{\mathbb{C}}
\def\Z{\mathbb{Z}}
\def\R{\mathbb{R}}
\def\d{\partial}

\def\vp{\varphi}

\def\Tr{\operatorname{Tr}}
\def\<{\langle}
\def\>{\rangle}

\def\P{\mathbb{P}}
\def\ad{\operatorname{ad}}
\def\Sym{\operatorname{Sym}}

\title{Topological quiver matrix models and quantum foam}
\author{Daniel L. Jafferis {\footnote{E-mail: jafferis@string.harvard.edu}}\\
Jefferson Physical Laboratory\\Harvard University\\Cambridge, MA
02138}
\date{May 15, 2007}
\begin{document}

\maketitle

\abstract{We study the matrix models that describe the BPS bound
states of branes arising from the quiver picture of the derived
category. These theories have a topological partition function
that localizes to the Euler character of the anti-ghost bundle
over the classical BPS moduli space. We examine the effective
internal geometry of D6/D2 bound states in the local vertex
geometry, using BPS 0-brane probes. The Kahler blowups of the
Calabi-Yau that we find utilizing these quiver theories are a
realization of A-model quantum foam in the full IIA theory.}

 \pagebreak

\begin{section}{Introduction}

The idea that the topological A-model involves quantized
fluctuations of the Kahler geometry of a Calabi-Yau was introduced
in \cite{qfoam}. They argued that the topological string partition
function could be reproduced by summing over non-Calabi-Yau blow
ups along collections of curves and points. The equivalent
Donaldson-Thomas theory, given by a topologically twisted ${\cal
N}=2$ \ $U(1)$ gauge theory on the Calabi-Yau, involves a sum over
singular instantons, which can be blown up to obtain the
fluctuations of the geometry.

The connection between this theory of BPS bound states of D2 and
D0 branes to a 6-brane was further explained in \cite{DVV} by
lifting to M-theory. The D6-brane lifts to a Taub-NUT geometry in
11 dimensions,  and the Donaldson-Thomas theory results in
precisely that repackaging of the Gopakumar-Vafa invariants
counting bound states of D2/D0 at the center of the Taub-NUT which
is required to reproduce the A-model.

In this work, I will try to elucidate the role of Kahler quantum
geometry when the topological A-model is embedded in the full IIA
theory. We will show that the effective internal geometry
experienced by BPS 0-brane probes of bound states of 2-branes and
a D6 is exactly the blow up of the Calabi-Yau along the wrapped
curves! This will be done in the context of the quiver matrix
models that describe the low energy dynamics, so we will see the
topological string theory emerge from a matrix model, with quite a
different flavor than \cite{DV2}.

Therefore, if such a bound state in the mixed ensemble were
constructing in $\R^{3,1}$, then in the probe approximation, the
effective geometry of the Calabi-Yau would be literally the
``foam'' envisioning in \cite{qfoam}. Moreover, the work of
\cite{DM} relating D4/D2/D0 bound states with classical black hole
horizons to multi-centered D6/$\overline{D6}$ configurations means
this might have implications for BPS black holes with
non-vanishing classical entropy.

More subtle probes of the geometry using $N$ 0-branes are able to
discern the presence of a line bundle associated to the
exceptional divisors of the blow up. Thus we see that the 2-branes
are blown up to 4-branes, which can be dissolved into $U(1)$ flux
on the D6. Moreover, only a small number of 0-branes can ``fit''
comfortably on the exceptional divisors, which might indicate that
their Kahler size is of order $g_s$, as predicted in \cite{qfoam}.
We will exactly reproduce the know topological vertex amplitudes
from the quivers we discover, by counting fixed points of the
equivariant $T^3$ action.

The topological quiver matrix models we shall discuss are quite
interesting in their own right. Consider the BPS bound states of
D6/D4/D2/D0 system in IIA theory compactified on a Calabi-Yau
3-fold. For large Kahler moduli, the D6 brane worldvolume theory
is described by the 6+1 dimensional nonabelian Born-Infeld action.
At low energies, we can integrate out the Kaluza-Klein modes in
the internal manifold, presumably giving rise to a 0+1 dimensional
superconformal quantum mechanics, which is the gauge theory dual
to the $AdS_2$ near horizon geometry. The full superconformal
theory remains mysterious, but there have been numerous attempts
to understand it, see for instance \cite{SCQM}.

Our interest is in the index of BPS ground states, which is
captured by the topological twisted version of the physical
theory. In the case of D6 branes, this is the twisted ${\cal N} =
2$ Yang Mills in six dimensions. The low energy description is now
a 0-dimensional topological theory, that is a matrix model with a
BRST trivial action. This topological version of the dual theory
is far simpler, and can be written down exactly in non-trivial
situations.

%For quiver matrix models describing the bound states of D4/D2/D0
%branes, it would be interesting to relate this to the MSW CFT
%obtained by lifting to M5 branes in M-theory. The IR limit of the
%M5 brane lives on the torus defined by time cross the M-circle.
%The target space of the resulting conformal theory is the moduli
%space of deformations of the 4-cycle wrapped by the M5. In some
%cases, the elliptic genus must reduce to a zero dimensional theory
%with a quiver description of the type we discuss here.

%The MSW description is most natural when the wrapped 4-cycle is
%very ample, with a large moduli space of deformations. On the
%other hand, the quiver description seems most natural in the
%opposite regime of rigid divisors, which may be related to the
%reduction of the theory to zero dimensions. In the CFT side these
%situations must correspond to having more constraints then fields.

The physical interpretation of the quiver matrix model changes
dramatically as one moves from the Gepner point to the large
volume limit in the Kahler structure moduli space. It is very
satisfying to see that the quiver description, which is most at
home at the Gepner point, gives the correct result for the index
of BPS states even in the opposite noncompact limit we will
consider. At large volume, the quiver should be thought of as the
collective coordinates of instantons in the worldvolume theory on
the D6 brane.

Note that the moduli on which the theory depends are really the
background moduli, defined at infinity in spacetime. In the case
of 6-brane bound state we are examining, there {\it are} no BPS
states at the naive attractor value of the moduli, so the
attractor flow is always a split flow. Amazingly, this doesn't
impair the ability of the quiver theory to capture the index of
bound states, and furthermore we are able to read off certain
aspects of the attractor flow tree \cite{DM} from the shape of the
matrix model potential.

One of the most powerful worldsheet techniques to explore string
theory compactified on a Calabi-Yau given by a complete
intersection in projective space are gauged linear sigma models
with superpotential. These two dimensional worldsheet theories can
be constructed explicitly and exactly, and flow in the IR to the
conformal non-linear sigma on the Calabi-Yau. For the purpose of
understanding BPS saturated questions, the linear model itself
already gives the exact answer, just as in the context of the
topological string; finding the IR fixed point is unnecessary.

The holomorphically wrapped D-branes we are interested in are
described by boundary linear sigma models, which have been studied
extensively in the context of the B-model. The branes of the
topological B-model can be encoded in the boundary conditions for
the open B-model worldsheet, which are naively associated to
vector bundles on the Calabi-Yau. In fact, a much richer structure
of boundary data has been discovered, associated to D6 and anti-D6
branes with tachyons condensed. Mathematically speaking, the
holomorphic branes are objects in the derived category of coherent
sheaves \cite{MRD}, with the tachyons becoming maps between the
sheaves associated to the 6-branes. The BPS branes of IIA theory
require a stability condition (for generic Kahler moduli,
$\pi$-stability of the triangulated category) not present for
branes of the topological B-model.

At the Gepner point in the Calabi-Yau moduli space, this
description simplifies, and one can generate all of the
configurations of branes by condensing tachyons between a complete
collection of fractional branes. The chain maps of the derived
category are then encoded in the linear maps between a quiver of
vector spaces, as shown in \cite{DGM}. In general there will be a
nontrivial superpotential, which must be minimized to obtain the
supersymmetric ground states \cite{aspthree}. The effective
topological theory of BPS branes is therefore the topological
quiver matrix model, whose partition function localizes to the
Euler characteristic of the anti-ghost bundle over the Kahler
quotient of the quiver variety.

As we move in the Calabi-Yau moduli space out to the large volume,
certain degrees of freedom will become frozen in the local limit.
We find the quiver for the remaining dynamical degrees of freedom,
and discover that there are a few extra terms in the effective
action. In the local limit, branes with different dimensions, or
those wrapping cycles with different asymptotics in the local
Calabi-Yau, become sharply distinguished. This corresponds to the
breaking of some gauge groups $U(N+M) \To U(N) \times U(M)$ of the
Gepner point quiver. There are residual terms in the action from
the off-diagonal components of the original D-term, which need to
retained even in the local limit. Crucially, however, the local
limit is universal, and the action we find is independent of the
specific global geometry into which is was embedded, so our
analysis is consistent.

The index of BPS states always looks, from the large volume
perspective, like the Euler character of some bundle over the
resolution of the instanton moduli space. The classical instanton
moduli are described by the zero modes of the instanton solutions
in the topologically twisted gauge theory living on the 6-branes.
We will find this moduli space more directly, using the quiver
description of the derived category of coherent sheaves, and
taking the appropriate large volume limit.

The resulting holomorphic moduli space literally gives the moduli
space of B-branes as the solution to some matrix equations
\eqref{quiver}, moded out by the gauge group. There are exactly
the pieces of the open string worldsheet gauged linear sigma model
which survive the topological string reduction to the finite
dimensional $Q$-cohomology. This moduli space is explicitly
independent of the Kahler moduli.

It is crucial that much of this moduli space does not correspond
to BPS bound states of holomorphically wrapped branes in IIA
theory. This is already clear because the stability of bound
states depends on the background Kahler moduli. There is a clear
realization of this fact in the gauge theory living on the branes,
whose instantons satisfy the Hermitian Yang-Mills equations
\eqref{hYM} \eqref{hYM2}. The holomorphic moduli space only
imposes the F-term conditions, namely \begin{equation} F^{2,0} =
0, \label{hYM}
\end{equation} and gauges the complexified gauge group, $GL(N,
\C)$, while the physical instanton moduli space is the Kahler
quotient, obtained from the D-term condition,
\begin{equation} F^{1,1} \wedge \omega^{d-1} = r \omega^d,
\label{hYM2}
\end{equation} which plays the role of the moment map, together
with the compact gauge group $U(N)$.

In the case of D4/D2/D0 bound states on local Calabi-Yau, whose
BPS sector is described by the Vafa-Witten twist of ${\cal N}=4$
Yang Mills theory on a 4-manifold, the partition function reduces
to the Euler character of the instanton moduli space. There the
quiver realization is exactly the ADHM construction of this finite
dimensional space. This can be promoted to a matrix model with the
same partition function, moreover the structure of the fermionic
terms will be associated to the tangent bundle over the classical
moduli space. Satisfyingly, this theory has exactly the 10
expected scalars and 16 fermions of the dimensional reduction of
the maximally supersymmetric gauge theories in higher dimensions.
Moreover, the action looks like a topologically twisted version of
the usual D0 worldvolume theory. In flat space, there are no
dynamical degrees of freedom associated to the motion of the
noncompact D2 and D4 branes, but there are near massless
bifundamentals associated to the 4-0 and 2-0 strings.

Our matrix models are the analogous construction for 6-branes.
There is no reason for the antighost bundle of the matrix model to
be the tangent bundle, although their dimensions are equal when
the virtual dimension of the moduli space is zero. In fact, we
will find that this obstruction bundle is indeed different in the
case of D6/D0 bound states.

The quivers we will find provide are sensitive probes of the
effective geometry, enabled one to go beyond calculations of the
Euler characteristic. An interesting structure emerges, in which
different effective quivers for the dynamical 0-brane fields are
related by flops in the blow up geometry of the Calabi-Yau they
probe. This leads to a more detailed understanding of the quantum
foam picture of the topological A-model.

\end{section}

\begin{section}{Review of quivers and topological matrix models}

\begin{subsection}{A quiver description of the classical D6-D0
moduli space}

We will begin by constructing the classical moduli space of $N$ D0
branes in the vertex geometry. The gauged linear model can be
obtained by dimensional reduction of the instanton equations in
higher dimensions. Consider bound states of holomorphically
wrapped branes which can be expressed as flux dissolved into the
worldvolume of the top dimensional brane. Then the classical
moduli space of these instantons is determined by the solutions of
the Hermitian Yang-Mills equations,
\begin{equation} \begin{split} F^{(2,0)} = 0
\\ F^{(1,1)} \wedge \omega^{d-1} = r \omega^d . \label{inst} \end{split} \end{equation}

The reduction of the 6d gauge field to zero dimensions results in
6 Hermitian scalar fields, which can be conveniently combined into
complex $Z_i$, for $i=1,2,3$. In terms of these matrices, the
F-term condition is the zero dimensional reduction of $F^{(2,0)} =
0$,
\begin{equation} [Z_i , Z_j] = 0 . \label{comm} \end{equation}
We find the D-term constraint as the analog of $F^{(1,1)} = r
\omega$, where $\omega$ is the Kahler form,
\begin{equation} \sum_{i=1}^3 [Z_i, Z_i^\dag] =0, \label{D0term}
\end{equation} which one can think of as the moment map associated
to the $U(N)$ action on adjoint fields. Here it is impossible to
find solutions with nonzero $r$, since the left hand side is
traceless. This is equivalent to the fact that one cannot add a
Fayet-Iliopoulos term when there are only adjoint fields. The
F-term condition \eqref{comm} comes from the superpotential
\begin{equation} W = \Tr Z_1 [Z_2, Z_3], \label{super0}
\end{equation} by the usual BPS condition that $\d W = 0$.

The gauged linear sigma model description of the topologically
twisted version of the worldvolume theory of $N$ D0 branes in
$\C^3$ can be obtained by computing the appropriate $Ext$ groups
in the derived category description. The fields in the quiver
theory are exactly the 0-0 strings that in this case give three
complex adjoint fields, $Z_i$, of $U(N)$, which can be naturally
interpreted as the collective coordinates of $N$ points in $\C^3$.

Adding the background D6 brane extends this quiver by a new $U(1)$
node, with a single field, $q$, coming from the 0-6 strings,
transforming in the bifundamental, as shown in \cite{Witten}. As
described by Witten, the new quiver data which determines the zero
dimensional analog of the instanton equations \eqref{inst} is
\begin{equation}
\begin{split} [Z_i, Z_j] = 0 \\ \sum_{i=1}^3 [Z_i, Z_i^\dag] +
q^\dag q = r I_N \label{quiver} \end{split} \end{equation} The
Fayet-Iliopoulos parameter, $r$, will turn out to be the mass of
the 0-6 fields, as we will see more clearly in the matrix model.
This mass is determined by the asymptotic B-field needed to
preserve supersymmetry in the D6/D0 bound states \cite{Witten}. In
the absence of a B-field, $r<0$, and $q$ is a massive field, hence
there would be a stable nonsupersymmetric vacuum. For $r>0$, which
is the case of interest for us, the now tachyonic $q$ condenses to
give a supersymmetric bound state. Note that there are no new
possible terms in the superpotential \eqref{super0} because of the
shape of the quiver, hence the bifundamental can only appear as we
have written it in the D-term.

The holomorphic quotient space that is naturally determined by the
F-term is ${\cal M}_{hol} = X / GL(N, \C)$, where $X$ is the space
of commuting complex matrices. This is exactly the moduli space of
branes in the B-model, which depends purely on holomorphic data.
It turns out to be a larger space, with less structure, than the
Kahler quotient ${\cal M}_{Kahler} = X // U(N)$ that describes the
physical BPS bound states in II theory. If an additional stability
condition, which in this case is the existence of a cyclic vector
for the representation of the quiver, is imposed on ${\cal
M}_{hol}$ then it can be shown to be identical to the Kahler
quotient, following the reasoning of \cite{OPN}.

\end{subsection}

\begin{subsection}{Quiver matrix model of D0 branes}

The index of BPS states of 0-branes can be found using the
topologically twisted reduction of the D0 matrix model of
\cite{Matrixtheory}, where the time directions drops out for the
obvious reasons when considering the structure of ground states.
This results in a zero dimensional theory, ie. a matrix model,
with a BRST like supersymmetry.

 Following \cite{MNS1} and
\cite{MNS2}, we will first consider the theory describing only D0
branes in $\C^3$, by utilizing the formalism developed by
\cite{VW} to study Euler characteristics by means of path
integrals. The action of the topological supersymmetry is given by
\begin{equation*}
\begin{split} \delta
Z_i = \psi_i , \ \delta \psi_i = [\phi, Z_i], \ \delta \phi = 0 \\
\delta \bar\phi = \eta, \ \delta \eta = [\phi, \bar\phi] \\
\delta \varphi = \zeta, \ \delta \zeta = [\phi, \varphi] \\
\delta \chi_i = H_i, \ \delta H_i = [\phi, \chi_i] \\
\delta \chi = H, \ \delta H = [\phi, \chi], \end{split}
\end{equation*} where all fields are complex $U(N)$ adjoints,
except for $\chi$ and $H$ which are Hermitian. This corresponds to
the $d=10$ case of \cite{MNS2} in a slightly different notation
that is more convenient in the context of compactification on a
Calabi-Yau 3-fold. The $Z_i$ are the D0 collective coordinates in
the internal manifold, while $\varphi$ morally lives in the bundle
of $(3,0) + (0,3)$ forms over the Calabi-Yau, which practically
means that it is a scalar.

This is the reduction to zero dimensions of the twisted ${\cal N}
= 4$ Vafa-Witten theory, and equivalently, of the twisted ${\cal
N}=2$ gauge theory in six dimensions. We will write a $Q$-trivial
action for this topological matrix model, which can be interpreted
as the topological twisted version of the $0+1$ dimensional
superconformal quiver quantum mechanics \cite{SCQM} describing
these bound states. The topological property results from the fact
we are only interested in computed the index of BPS states, which
can be described in this zero dimensional theory.

Choose the sections, in the language of \cite{VW}, to be
\begin{equation}
\begin{split} s_i = \Omega_{i j k} \ Z_j Z_k + [\varphi, Z_k^\dag]
\\ s= \sum_i [Z_i, Z_i^\dag] + [\varphi^\dag, \varphi],
\label{sections}
\end{split} \end{equation} where $\Omega_{i j k}$ is the
holomorphic 3-form, which we can assume to be given by the
antisymmetric $\epsilon_{ijk}$ for the $\C^3$ vertex geometry.

The action is given by $S = t \{Q, V \}$, where \begin{equation*}
V = \Tr \left( \chi_i^\dag (H_i - s_i) \right) + \Tr \left( \chi
(H-s) \right) + \Tr \left(\psi_i [\bar\phi, Z_i^\dag] \right) +
\Tr \left(\zeta^\dag [\bar\phi, \varphi] \right) + \Tr \left( \eta
[\phi, \bar\phi] \right).
\end{equation*} Following \cite{VW}, we write the bosonic terms
after integrating out $H_i$ and $H$, which appear quadratically in
the action, to obtain
\begin{equation} \label{boson} S_{bosonic} = \Tr \left( s_i^\dag s_i + s^2 +
[\phi, Z_i] [\phi, Z_i]^\dag + [\phi, \varphi^\dag][\phi,
\varphi^\dag]^\dag +[\phi, \bar\phi]^2 \right),
\end{equation} where the $\Tr [\phi, Z_i][\phi, Z_i]^\dag$ terms
arise from the twisted superpotential, thus coupling the vector
multiplet, $\phi$, with the chiral matter fields charged under it.

This theory has a total of 9 real scalars, coming from the $Z_i$,
$\varphi$, and $\phi$, transforming the adjoint of $U(N)$,
corresponding to the 9 collective coordinates of a point-like
brane. It is clear, following the reasoning of \cite{VW}, that the
partition function will count, with signs, the Euler character of
the moduli space of gauge equivalence classes of solutions to the
equations, $s_i = 0, \ s=0$. Moreover, the field $\phi$ can be
easily eliminated from the theory, since it simple acts to enforce
the $U(N)$ gauge symmetry on the moduli space. To relate this to
the ADHM type data we discussed in the previous section, we need
to understand what happens to the extra collective coordinate,
$\varphi$. In fact, we will prove a vanishing theorem, and find
that it does not contribute to the instanton moduli space. Its
main function in the matrix model is to ensure that all of the
solutions are counted with the same sign, so that the Euler
character is correctly obtained, in the same spirit as \cite{VW}.

Note that there are 4 dynamical complex adjoint fields in the
matrix model, the $Z_i$ and $\varphi$, which satisfy 3 holomorphic
equations \eqref{sections}. The Kahler quotient by $U(N)$ gives 1
real D-term equation from the moment map, and quotients by $U(N)$.
Hence the expected dimension of the moduli space is $8-6-1-1 = 0$,
however it may not consist of isolated points in practice unless
the sections are deformed in a sufficiently generic manner.

In order to fully make sense of this theory, we need to regulate
the noncompactness of $\C^3$ in some way. For local analysis on
toric Calabi-Yau, the natural choice is to add torus-twisted mass
terms, which will further enable one to localize the path integral
to the fixed points of the induced action of $U(1)^3$. It is
possible to understand the toric localization in the matrix model
language. We want to turn on the zero dimensional analog of the
$\Omega$ - background explained in \cite{qfoam}. Exactly as in
that case, we twist the $U(N)$ gauge group by the torus $U(1)^3$
symmetry. That is, we gauge a nontrivial $U(N) = SU(N) \times
U(1)$ subgroup of $U(N) \times U(1)^3$.

The effect of this on the topological matrix model is to change
the BRST operator so that $Q^2$ generates one of the new gauge
transformations. That is, the fields transform as \begin{equation}
\begin{split} \delta Z_i = \psi_i \\ \delta \psi_i = [\phi, Z_i] -
\epsilon_i Z_i, \end{split} \end{equation} where the $\epsilon_i$
determine the embedding of the gauged $U(1)$ inside $U(1) \times
U(1)^3$.

This changes the terms involving $\phi$ in the bosonic action to
become $\sum \Tr ( [\phi, Z_i] - \epsilon_i Z_i ) ([Z_i^\dag,
\bar\phi] - \epsilon Z_i^\dag) $, which forces the adjoint fields
$Z_i$ to be localized with respect to the torus action. We will
soon see that the partition function is independent of the choice
of weights $\epsilon_i$ as long as the superpotential itself is
invariant under the chosen group action, so that \begin{equation*}
\epsilon_1+\epsilon_2 +\epsilon_3=0.\end{equation*} This is
natural, since only these torus actions are subgroups of the
$SU(3)$ holonomy, and thus the equivariant twisted theory
continues to preserve supersymmetry.

There is an induced action of $U(1)^3$ on the other fields in the
theory, $\chi_i \To e^{i \alpha - i \epsilon_i} \chi_i$ and
$\varphi \To e^{î\alpha} \varphi$, where $\alpha = \sum
\epsilon_i$, which changes the action of $Q$ in the obvious way.

To understand the measure factor at the solutions that contribute
to the partition function we need to examine the fermionic piece
of the action, \begin{equation} \begin{split} S_f = \Tr \left(
\phi [\chi_i, \chi_i^\dag]+\bar\phi ( [\psi_i, \psi_i^\dag] -
[\zeta, \zeta^\dag] )+  \Omega_{i j k} \psi_i[  Z_j, \chi_k^\dag]
+
 \zeta [Z_i^\dag, \chi_i^\dag] - \psi_i^\dag [\chi_i^\dag ,\vp]
 \right)\\
 +\Tr \left( (\chi+i\eta) ([\psi_i, Z_i^\dag]-[\zeta, \vp^\dag])  \right).
 \end{split} \end{equation} The anti-ghost, $\chi$, which enforces the moment
 map
condition, naturally pairs with $\eta$, the fermionic partner of
the $U(N)$ multiplet, to produce the Kahler quotient of the moduli
space. We see that $\eta$ appears linearly in the action, thus
when it is integrated out, the delta function constraint $[\psi_i,
Z_i^\dag]-[\zeta, \vp^\dag]=0$ is enforced. This forces the
fermions $\psi_i$ and $\zeta$ to lie in the sub-bundle of the
tangent bundle normal to the D-term condition, $s=0$. Therefore
the Euler character of the Kahler quotient will be obtained, as in
\cite{VW}.

The topological nature of the partition function means that we are
free to change the coupling, $t$, which results in a BRST trivial
change of the theory, leaving the partition function invariant.
Taking the limit $t \rightarrow \infty$, the matrix model
localizes to the classical moduli space defined by
\begin{equation} \begin{split} \Omega_{i j k} Z_j Z_k + [\varphi,
Z_i^\dag] = 0 \\ \sum_i [Z_i, Z_i^\dag] + [\varphi^\dag, \varphi]
= 0 ,
\end{split} \end{equation} where we must mod out by the gauge
group, $U(N)$.

Near the solutions to the BPS equations the potential can be
approximated as a Gaussian, and the partition function will pick
up a determinant factor. From the bosonic part of the action
\eqref{boson} expanded near a solution, there are quadratic terms
of the form $\Tr(\hat\delta s_i^\dag \hat\delta s_i)$, where here
$\hat\delta$ is the variation. Terms such as $\Tr(s_i^\dag
\hat\delta (\hat\delta s_i))$ cannot arise since $s_i = 0$ for the
BPS solutions. For each such contribution, there is an analogous
fermionic term of the form $\Tr(\chi_i^\dag \delta s_i)$. After
integrating out the anti-ghosts, the quadratic terms involving the
ghosts $\psi_i$ will exactly match those of the fields $Z_i$. Call
the resulting quadratic form ${\cal A}$.

In addition there are the twisted superpotential terms of the form
$\left| [\phi, Z_i]+\epsilon_i Z_i \right|^2$, which give a total
contribution to the one loop determinant \begin{equation}
\frac{1}{\det( \ad \phi + \alpha)} \frac{\det (\ad \phi +\alpha -
\epsilon_i)}{\det (\ad \phi + \epsilon_i)}, \end{equation} which
exactly cancels, up to sign when we include the measure factor for
diagonalizing $\phi$, if $\alpha = \sum \epsilon_i = 0$, as in
\cite{MNS2}. This is all there is for the bound state of $N$
0-branes, since there are no nontrivial fixed points, hence ${\cal
A}=0$. For future reference, the answer in general would be
\begin{equation} \frac{\det(A+T)}{\det(A+T')}, \end{equation}
where $T$ and $T'$ are the $\epsilon_i$ dependent pieces in the
quadratic approximation. Thus for $\sum \epsilon_i = 0$, we still
have exact cancellation, even at the nontrivial fixed points. This
is no surprise, since we designed the theory for precisely this
effect.

Naively one would expect that the anti-ghost bundle whose Euler
character the matrix model will compute is simply spanned by
$\chi_i$ and $\chi$ fibered trivially over the moduli space.
However, by looking at the fermion kinetic terms involving
$\psi_i$ and $\zeta$, one finds that the bundle is obstructed when
restricted to the manifold obtained after imposing the F-flatness
condition. In particular, there are terms in the action,
\begin{equation} \Tr \left( \psi_i \left(\Omega_{i j k}
[\chi_j^\dag, Z_k] - [\chi, Z_i^\dag] \right) + \zeta \left(
[\chi_i^\dag, Z_i^\dag] + [\chi, \vp^\dag] \right) \right),
\end{equation} which require that \begin{equation} \begin{split} \Omega_{i j k}
[Z_i^\dag, \chi_j] = [Z_k, \chi] \\ [Z_i, \chi_i] = [\chi, \vp].
\end{split} \end{equation} This defines the anti-ghost bundle, or rather
complex of bundles, whose Euler character is the index of BPS
states computed by the matrix integral. It is obviously distinct
from the tangent bundle, although it has the same rank, and their
Euler characters may thus differ.

Now we proceed to find the vanishing theorem that shows the field
$\vp$ does not contribute to the instanton equations. First,
notice that using the F-term condition, \begin{equation} \sum_k
|[\varphi^\dag, Z_k]|^2 = \Tr[\varphi^\dag, Z_k] \Omega_{ijk} Z_j
Z_k = \frac{1}{2} \Tr \varphi^\dag \Omega_{ijk} [Z_k, [Z_j, Z_k] ]
= 0,
\end{equation} by the Jacobi identity, and the antisymmetry of
$\Omega_{ijk}$. This means that \begin{equation} \label{pz}
[\varphi^\dag, Z_k] = 0, \end{equation} and moreover $[Z_i, Z_j] =
0$, as we had hoped.

What can we learn from the D-term? Rewriting it as
\begin{equation} 0 = \left| \sum [Z_i , Z_i^\dag] \right|^2 + \left| [\varphi^\dag, \varphi] \right|^2 + 2 \sum \Tr
[\varphi^\dag, \varphi] [Z_i, Z_i^\dag] , \end{equation} we see
that all the terms are positive, since \begin{equation} \Tr
\varphi^\dag \left[\varphi, [Z_i, Z_i^\dag] \right] = - \Tr
[\varphi^\dag, Z_i] [Z_i^\dag, \varphi] - \Tr [\varphi^\dag,
Z_i^\dag] [\varphi, Z_i] = \left| [Z_i, \varphi] \right|^2,
\end{equation} by the Jacobi identity and equation \eqref{pz}.
Therefore we see that $\varphi$ commutes with both the $Z_i$ and
their conjugates, and can be trivially factored out of the theory.

At this point, all of the fields are on the same footing, since we
can use the same reasoning to show that $[Z_i, Z^\dag_j] = 0$, and
the theory localizes to the trivial branch of moduli space. This
is exactly the marginally supersymmetric state of $N$ independent
D0 branes. Note that if we rewrite the adjoint fields in terms of
9 Hermitian collective coordinates, and expand all of the terms
appearing in the action, we will see the full $SO(9)$ rotational
symmetry of non-relativistic D0 branes in flat space. This will
soon be broken by the addition of a D6 brane, and the presence of
a B-field.

\end{subsection}

\end{section}

\begin{section}{Solving the matrix model for D6/D0 in the vertex}

Now we will proceed to find the matrix model description of the
quantum foam theory of bound states of 1 D6 brane and $N$ D0
branes in the vertex. The only new field comes from the 0-6
strings, which results in a chiral scalar in the low energy
description. This appears in the matrix model as a new topological
multiplet, \begin{equation} \begin{split} \delta q = \rho \\
\delta \rho = \phi q, \end{split} \end{equation} living in the
fundamental of $U(N)$. See the left side of Figure \ref{vq} for
the quiver diagram of the internal Calabi-Yau degrees of freedom.

It is clear that moving the D0 branes off the D6 brane in the
normal directions will give a mass to the 0-6 strings. Thus there
should be a term in the quiver action of the form $q^\dag \vp
\vp^\dag q$. The similar mass term, $q^\dag \bar\phi \phi q$, is
automatically included in the twisted superpotential, as we shall
see below. Moreover, the BPS bound states must therefore have
$\vp^\dag q = 0$, which, however cannot obviously be implemented
by any superpotential. This requires the addition of new
anti-ghosts,
\begin{equation}
\begin{split} \delta \xi = h
\\ \delta h = \phi \xi , \end{split} \end{equation} living in the
fundamental of $U(N)$. The number of new fields is thus cancelled
by the new equations, and we are still left with a moduli space of
expected dimension 0.

As we will see later in similar examples, this extra equation
should be understood as a symptom of working in a noncompact
geometry. Quiver theories normally arise by considering the theory
of fractional branes at a Gepner point where the central charges
are all aligned. The exact nature of that quiver depends upon the
compactification, but in the local limit, will be reduced to those
found here. The general pattern is that a larger gauge group gets
broken to $U(N) \times U(M)$ where the $N$ and $M$ are the numbers
of branes that are distinguished (for example as D0 and D6) in the
large volume limit we are interested in, far from the Gepner
point. There will typically be some off-diagonal D-term which
persists even in the large volume limit, and $\vp^\dag q = 0$ is
just such an example.

\begin{figure}
\begin{center}
$\begin{array}{c@{\hspace{1in}}c}
\epsfig{file=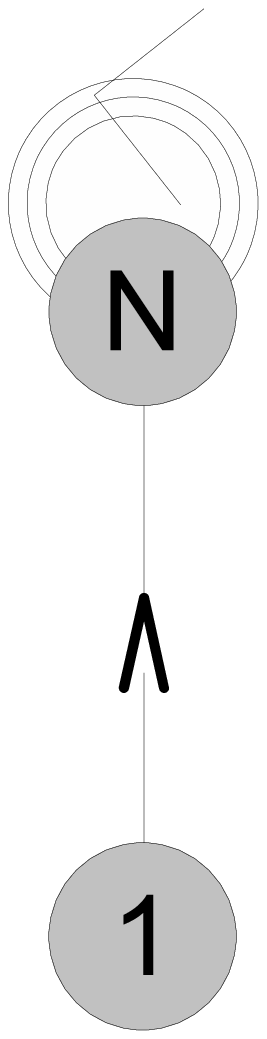,height=5cm} &
\epsfig{file=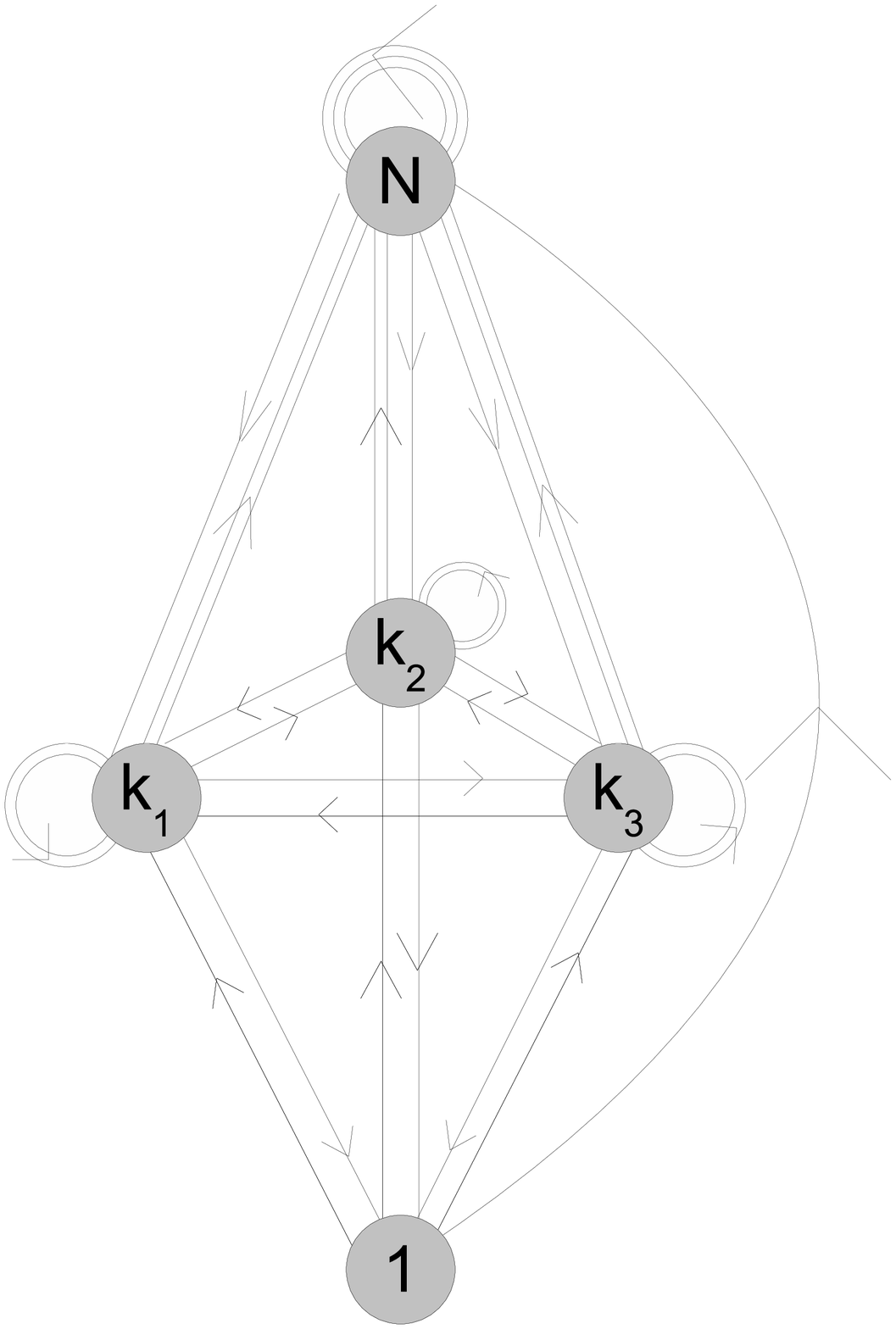,width=5cm} \end{array}$ \caption{The
quiver diagrams for the topological vertex with trivial and
generic asymptotics.} \label{vq}
\end{center}
\end{figure}

The $U(N)$ D-term now includes the contribution of the
bifundamental, replacing the section $s$ by \begin{equation} s =
\sum_i [Z_i, Z_i^\dag] + [\varphi^\dag, \varphi] + q q^\dag - r I,
\end{equation} where $r$ is the mass of the field $q$. The
anti-ghost $\xi$ appears in the action in the form
\begin{equation} \{Q, \xi^\dag (h -  \varphi q) \}. \end{equation}

Putting everything together, we find that the bosonic part of the
action is \begin{equation} \begin{split} S_b =  \left|
\Omega_{ijk} Z_i Z_j + [\varphi, Z_k^\dag] \right|^2 + \left|
\sum_i [Z_i, Z_i^\dag] + [\varphi^\dag, \varphi] + q q^\dag  - r I
\right|^2 \\ + \ q^\dag \varphi \varphi^\dag q+ \left| [\phi, Z_i]
\right|^2 + \left| [\phi, \varphi^\dag] \right|^2 + \Tr [\phi,
\bar\phi]^2 .
\end{split}
\end{equation} It is clear that the only effect of the Fayet-Iliopoulos parameter, $r$,
is to make the 0-6 strings tachyonic by adding the term $-r \ q
q^\dag$ to the action. As explained in \cite{Witten} the mass of
these bifundamentals is determined by the asymptotic B-field. In
particular, they are massive, $r<0$, in the absence of a magnetic
field, so that the minimum of the action has nonzero energy and
there are no BPS states. For a sufficiently strong B-field, $q$
will become tachyonic and can condense into a zero energy BPS
configuration.

The argument we used before implies that the theory localizes on
solutions to the equations,
\begin{equation}
\begin{split} [Z_i, Z_j] = 0 \\ [Z_i, \varphi^\dag] = 0 \\ \sum_i [Z_i,
Z_i^\dag] + [\varphi^\dag, \varphi] + q^\dag q = r I \\ q^\dag
\varphi = 0.
\end{split} \end{equation} Note that equation \eqref{pz} still
holds, since the superpotential it was derived from is unchanged.

We find a vanishing theorem from the D-term as before, computing
\begin{equation}
\begin{split} \label{Dterm} 0 = \left| \sum_i [Z_i, Z_i^\dag] + [\varphi^\dag,
\varphi] + q^\dag q - r I \right|^2 = \left| [Z_i, Z_i^\dag] +
q^\dag q - r I \right|^2 + \Tr [\varphi^\dag, \varphi]^2 + \\ 2
\sum \Tr[\varphi^\dag, \varphi] [Z_i, Z_i^\dag] + 2 q^\dag
[\varphi^\dag, \varphi] q .
\end{split}
\end{equation}

Only the last term is different, and using the fact that $ q^\dag
\varphi = 0$, as required by the potential for moving the D0
branes off the D6 in the spacetime direction, we see that all
terms on the right hand side of \eqref{Dterm} are positive
squares. Hence the moduli space consists of solutions to the
quiver equations \eqref{quiver}, with the field $\varphi$ totally
decoupling from the classical moduli space. It will, however, give
a crucial contribution to the 1-loop determinant at the fixed
points.

\begin{subsection}{Computing the Euler character of the classical
moduli space}

The moduli space of BPS states is given by the solutions to
\eqref{quiver} up to $U(N)$ gauge equivalence. We want to
determine the Euler characteristic of this quiver moduli space
using equivariant techniques. There is a natural action of the
$U(1)^3$ symmetry group of the vertex geometry on this classical
instanton moduli space, given by $Z_i \rightarrow \lambda_i Z_i, \
q \rightarrow \lambda q$. The Euler character is localized to the
fix points, which are characterized by the ability to undo the
toric rotation by a gauge transformation, that
is, \begin{equation} \begin{split} [\phi, Z_i] = \epsilon_i Z_i \\
q \phi - \alpha q= \epsilon q , \label{toric} \end{split}
\end{equation} for some $\phi \in su(N)$, and where $\alpha \in u(1)$ implements the $U(1)$ gauge
transformation, which can obviously be re-absorbed into the $U(N)$
transformation parameterized by $\phi$.

This is implemented in the matrix model as before, by changing the
${\cal Q}$ action so that it squares to the new $U(N)$ subgroup of
$U(N) \times U(1)^3$ that is being gauged. Consistency requires
that $\vp$ and $\xi$ have weights  $\epsilon_1+
\epsilon_2+\epsilon_3$ and $\epsilon - \epsilon_1+
\epsilon_2+\epsilon_3,$ respectively, under the torus action. The
twisted superpotential terms in the path integral are modified in
the obvious way, schematically, from $\Tr[\phi, \cdot] [\phi,
\cdot]^\dag$ to $\left| [\phi, \cdot] + \epsilon \ \cdot
\right|^2$.

The weight at the fixed point can be determined as follows,
assuming that $\epsilon_1+ \epsilon_2+\epsilon_3=0$ to preserve
the $SU(3)$ holonomy. It is easy to check as before that
contributions to the quadratic terms in the variation of $Z_i$
away from the fixed point exactly match those of $\psi_i$. The
twisted superpotential gives rise to additional contributions
\begin{equation} \Tr \left( [ \delta Z_i, \phi] + \epsilon_i \delta Z_i \right) \left( [\bar\phi, \delta
Z_i^\dag] + \epsilon_i \delta Z_i^\dag \right) + \left| [\vp,
\phi] \right|^2 + q^\dag (\bar\phi + \epsilon) (\phi + \epsilon)
q,
\end{equation} to the bosonic fields, and for the anti-ghosts,
\begin{equation} \left| [\chi_i, \phi] - \epsilon_i \chi_i
\right|^2 + \xi^\dag (\bar\phi + \epsilon) (\phi + \epsilon) \xi,
\end{equation} where we used the fact that our torus action lives in $SU(3)$.
Therefore, everything cancels exactly after including the
Vandermonde determinant from diagonalizing $\phi$.

To begin finding the fixed points, chose the gauge by requiring
that $\phi$ of \eqref{toric} is diagonal, denoting the eigenvalues
by $\phi_a$. This is a far more judicious choice then trying to
diagonalize the $Z_i$, since they are not Hermitian, and all of
the interesting bound states in fact require them to be
non-diagonalizable. The noncommutivity of $Z_i$ and $Z_i^\dag$
thus implied can be understood physically as resulting from the
background B-field we have turned on to produce supersymmetric
bound states in the D6/D0 system. This field is indeed
proportional to the Kahler form, $\omega = \sum_i d z_i^* \wedge d
z_i$.

In this gauge, the equivariant condition gives the strong
constraint that \begin{equation} (Z_i)_{a b} (\phi_a - \phi_b -
\epsilon_i) = 0, \ q_a (\phi_a - \alpha - \epsilon) = 0 .
\end{equation} For generic $\epsilon_i$, this forces most of the
components of $Z_i$ to vanish, and it is useful to think of the
nonzero elements as directed lines connecting a pair of the $N$
points indexed by $\phi_a$.

It is convenient to represent the action of the $Z_i$ as
translation operators on a finite collection, $\eta$, of $N$
points in $\Z^3$ associated to the eigenvectors of $\phi$, with
adjacency determined by the condition \begin{equation}\phi_a -
\phi_b = \epsilon_i, \label{tor}\end{equation} which must hold
whenever $(Z_i)_{ab} \neq 0$. More precisely, associating each
eigenvector of $\phi$ to a point $x^a$ in $\Z^3$, whenever
\eqref{tor} is satisfied, the points are related spatially by
\begin{equation*} x^a_i = x^b_i + 1. \end{equation*}In relating the
quiver description to the six dimensional Donaldson-Thomas theory,
recall that the fixed points of the induced $(\C^\times)^3$ action
on equivariant ideals of the algebra of polynomials on $\C^3$ can
be encoded as three dimensional partitions in the positive octant
in the lattice $\Z^3$ \cite{qfoam}. These are equivalent in the
standard way to coherent, torus invariant sheaves with point-like
support at the origin in $\C^3$.

What does the F-flatness condition, $[Z_i,Z_j]=0$, mean for the
torus localized configurations? Clearly this implies the fact that
if $p, q \in \eta$ can be connected by a particular sequence of
positive translations, $Z_i$, remaining within $\eta$ at each
stage, then all such paths generated by other orders of the same
$Z_i$ must also lie in $\eta$. It is easy to see that this means
that $\eta = \pi - \pi'$ is the difference between two three
dimensional partitions, $\pi' \subseteq \pi$. Moreover, the values
of the nonzero $(Z_i)_{a b}$ can be chosen such that the matrices
indeed commute.

This has a nice interpretation in the language of the derived
category of coherent sheaves used to describe all equivariant
B-model boundary states in the vertex geometry. Let ${\cal E}_\pi$
be the equivariant sheaf associated to the partition $\pi$
following \cite{qfoam}. Then the configuration for general $\eta$
is associated to the complex
\begin{equation*} {\cal E}_\pi \rightarrow {\cal E}_{\pi'},
\end{equation*} which, although perfectly fine as a B-model brane,
is always an unstable object in the full theory, for any value of
the B-field. It has D0 charge of $N$, and vanishing D6 charge,
since we have condensed the $D6/ \overline{D6}$ tachyon. At the
level of modules, the worldvolume of this B-brane is described by
$M = {\cal I}_{\pi'} / {\cal I}_{\pi}$. This is exactly what we
should have expected, since we haven't yet imposed the D-term
constraint, which gives the stability condition in the quiver
language.

The dimensional reduction of the $(1,1)$ part of the field
strength that appears in \eqref{quiver} is naturally diagonal in
the basis we have chosen. For most points in $\eta$ it is possible
to find $Z_i$ satisfying all the conditions, such that $[Z_i,
Z_i^\dag] = Z_i Z_i^\dag - Z_i^\dag Z_i$ has a positive eigenvalue
at that point. The exception are the points $p \in \eta$ on the
interior boundary, $\pi'$, since they are killed by $Z_i^\dag$ and
have only the negative contribution $- Z_i^\dag Z_i$.

Therefore to satisfy the D-term condition, these negative
eigenvalues must be cancelled by the contribution of the D6/D0
bifundamental, $q^\dag q$. The single D6 brane only gives us one
vector $q^\dag \in \C^N$ to work with, hence only one negative
eigenvalue of $[Z_i, Z_i^\dag]$ can be cancelled. Hence there are
only solutions when the interior has a single corner, namely when
$\pi'$ is trivial, and $\eta$ is a three dimensional partition.
Therefore we have reproduced the crystals first related to the
A-model in \cite{topcrystal}

In this case, the $Z_i$ act as multiplication operators in the
algebra ${\cal A} = \C[x,y,z] / {\cal I}_\pi$, which is the
algebra of polynomials on the nonreduced subscheme found in the
nonabelian branch of $N$ points in three dimensions. The
holomorphic version of the stability condition is the existence of
a cyclic vector, $q \in \C^N$, such that polynomials in the $Z_i$,
acting on $q$, generate the entire vector space. This is obvious
for $Z_i$ given by translation operators in the dimension $N$
unital algebra, ${\cal A}$, with $q = 1 \in \C[x,y,z]$. Moreover,
it can be shown that the D-term constraint is equivalent to this
algebraic stability requirement.

\end{subsection}

\begin{subsection}{Generalization to the $U(M)$ Donaldson-Thomas theory}

The $U(M)$ Donaldson-Thomas theory describes the bound states of
$M$ D6 branes with D2 and D0. The instanton moduli space is
described by the same quiver equations \eqref{quiver} as before,
where $q$ is now a $(N,\bar M)$ bifundamental.

Applying the analysis of section 2, we find that the $Z_i$ act as
translations on a nested partition, $\eta = \pi - \pi'$, with $N$
boxes. In the same way as before the moment map constraint can be
saturated only when $q^\dag q$ can cancel the negative
contributions to $\sum [Z_i, Z_i^\dag]$ on the interior corners.
Therefore $\eta$ can have at most $M$ interior corners.
Equivalently, choose $M$ points in $\Z^3$, and construct
overlapping three dimensional partitions based on each point. This
will define a permissible $\eta$.

Consider the example of $U(2)$ Donaldson-Thomas invariants. Then
it is clear from the quiver description of the moduli space that
the equivariant bound states are associated to partitions in an
``L-shaped'' background, as shown. This background is the most
general with two corners, moreover configurations consisting of
decoupled partitions resting independently on the corners are
obviously not included. Therefore we can explicitly determine the
partition function using the results of \cite{topcrystal} and
\cite{J} about the statistical mechanics of melting crystals,
obtaining
\begin{equation} \begin{split} Z_{U(2)} = \sum_{n,m,k > 0} \left(M(q) \ C_{[n m] [k n] \cdot}
\ q^{(n m^2 + n^2 k)/2} - S^{(n)} S^{(m, \ k)} \right) +
\\ \sum_{n,m>0} \left(M(q) \ C_{[n m] \cdot \cdot} \ q^{n m^2 / 2} -
S^{(n)} S^{(m)} \right) , \label{DT2} \end{split} \end{equation}
where $[n m]$ is the rectangular $n \times m$ Young diagram.

The first term is the generating function of crystals on the
L-shape given by the asymptotic rectangular Young diagrams, where
the power of $q$ is present because of the framing factor in the
topological vertex relative to the crystal partition function.
This term leads to an over-counting of contributions to the $U(2)$
theory because of the inclusion of decoupled partitions supported
independently in the two corners, hence these are subtracted off
by the second term. It is also possible for the two interior
corners to lay in the same plane, which is captured by the final
two terms in \eqref{DT2}. The partition functions $S^{(n, \ m)}$
are defined in \cite{J}.

Although the result \eqref{DT2} is not very transparent, it is
notable as the first calculation of the $U(2)$ Donaldson-Thomas
theory, even in the vertex. It would be very interesting to try to
confirm this formula mathematically, as well as the more implicit
(although still fully calculable order by order) answer for
$U(M)$.

\end{subsection}

\end{section}

\begin{section}{The full vertex $C_{\mu \nu \eta}$}

The topological A-model partition function has been shown to be
given by a dual description in terms of the bound states of a
single 6-brane with chemical potentials turned on for D2 and D0
branes at large values of the background B-field \cite{qfoam}. In
toric Calabi-Yau this has been checked explicitly, since both the
A-model and the Donaldson-Thomas theory localize onto equivariant
contributions in the vertex glued along the legs of the toric
diagram. The general vertex, which thus determines the entire
partition function of the A-model on any toric Calabi-Yau, can
also be described by a quiver matrix model.

We will construct this quiver, and see that it reproduces the
known answer for the topological partition function. In addition,
this model gives much additional interesting information about the
moduli space of these bound states then merely the Euler
character. We will find that in general, the effective geometry
seen by the dynamical D0 branes depends in an intriguing way on
the background Kahler moduli, and undergoes flop transitions as
one crosses walls where the attractor trees change shape, as in
\cite{DM}. In the next section, we will see how the quantum foam
picture of fluctuating Kahler geometry arises naturally in the
quiver description, and further surprises of the effective
geometry will become apparent.

The effective geometry we explore with 0-branes is, of course, the
geometry of the moduli space of BPS states. This can receive
various corrections both in $\alpha'$ and $g_s$. In the local
case, however, the situation is more under control, since the
Calabi-Yau moduli are always at large volume. Moreover, the torus
action extends to the effective geometry, and we see no reason
that this $T^3$ symmetry should be violated by corrections, at
least in the local limit. Hence the corrections to the Kahler
structure on the effective moduli appear likely to be simple
renormalizations of the already present Kahler parameters, ie. the
dependence of the FI terms on the background moduli receives
$\alpha'$ corrections, which is no surprise.

\begin{subsection}{One nontrivial asymptotic of bound D2 branes}

We want to determine the low energy effective quantum mechanics of
BPS D6/D2/D0 bound states. Note that the D6/D2 system is T-dual to
a 4-brane with bound 0-branes, and is described by the same ADHM
quiver \cite{ADHM}. As one dials the background values of the
Kahler moduli, the D6/D2/D0 configurations are more naturally
regarded as either point-like instantons in a D6 brane with D2
(singularly supported) ``flux'' turned on, a collection of
2-branes bound to a D6 with D0 charge, or a D2/D0 bound state
attached to the D6. Only the first interpretation will be relevant
for us, as we will soon see.

The low energy spectrum of 2-0 strings in flat space consists of a
single tachyonic bifundamental multiplet, denoted here by $B$,
which remains tachyonic for all values of the background B-field.
The BPS states are obtained by condensing this field to cause the
D0 branes to dissolve into flux on the D2. In our situation,
however, if we turn on a B-field, the 6-2 strings can become far
more tachyonic. As we will see quantitatively below, this field
condenses first, melting the D2 branes into $U(1)$ flux on the D6,
and giving a large positive contribution to the mass of the usual
D2/D0 and D6/D0 bifundamentals. They become irrelevant to the
moduli space of BPS states, but, surprisingly, an initially
massive multiplet in the 2-0 spectrum receives an opposite
contribution and becomes tachyonic.

A particularly simple way of determining the relevant spectrum of
low energy 2-0 fields motivated by string theory is to look at the
large volume limit of a D2-D0 bound state wrapping a compact
2-cycle. Consider $k$ D2 branes on the $S^2$ of the resolved
conifold, which are described by the quiver \cite{conifold} shown
in Figure ~\ref{cq}, with total D0 charge $N + k/2$ (including
induced charge). The superpotential implies that
\begin{equation} C_1 D_a C_2 = C_2 D_a C_1, \ \ \ \ D_1 C_a D_2 =
D_2 C_a D_1, \label{coni}
\end{equation} for the supersymmetric vacua.

\begin{figure}
\begin{center}
\epsfig{file=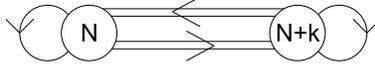,width=5cm} \caption{The quiver diagram
for the resolved conifold} \label{cq}
\end{center}
\end{figure}

 The motion
of the 0-branes is governed by the collective coordinates on the
conifold, $c_1 d_2 = c_2 d_1$, where $(d_1: d_2)$ are projective
coordinates on $\P^1$. At first, we want to focus on the case
where there are $N$ point-like bound D0 branes near one vertex. I
will work holomorphically to avoid considerations of stability,
which will be strongly affected by the D6 brane we will introduce
shortly . This can be done by fixing the gauge,
\begin{equation*} D_2 = \left(
\begin{array}{c c} I_N & 0_{N \times k} \end{array} \right),
\end{equation*} which implicity assumes that all of the D0 branes
are in the patch $d_2 \neq 0$, consistent with the desired
noncompact limit we will take to obtain the vertex geometry.

For this gauge choice, we have broken the $U(N) \times U(N+k)$
symmetry down to $U(N) \times U(k)$, and obtained the
``localized'' quiver shown. This gives the quiver for D2-D0 branes
in flat space, including the $A_a$ fields which are massive
without the presence of the D6 and background B-field. They will
play a crucial role for us, as we will soon see.

The remaining quiver fields for the conifold decompose under the
breaking of the gauge group as
\begin{equation} C_a =  \left( \begin{array}{c} Z_a \\ A_a \end{array} \right), \ \ \ \ D_1 =\left(
\begin{array}{c c} Z_3 & B \end{array} \right). \end{equation} The
F-flatness equation \eqref{coni} implies that \begin{equation}
[Z_1, Z_2] = 0, \ [Z_2, Z_3] = B A_1, \ [Z_3, Z_1] = B A_2, \ A_1
Z_2 = A_2 Z_1.
\end{equation} To obtain the full quiver for 2-0 in $\C^3$,
one needs to modify these equations in the natural way to include
the motion of the D2 branes, which are rigid in the conifold
example. These equations naturally result from extremizing the
superpotential \eqref{W026} below.

In order to obtain the correct moduli space for $N>1$ and $k>1$
these is one extra ingredient we have so far neglected. This can
be motivated in two different ways. First, in the above derivation
of the quiver by taking a limit of the resolved conifold geometry,
the D-term of the original $U(N+k)$ symmetry has an off-diagonal
component in the decomposition to $U(N) \times U(k)$. This would
imply the existence of an additional constraint, \begin{equation}
A_a Y_a^\dag + Z_a^\dag A_a = 0, \label{offdiag} \end{equation}
where we have included the generalization to a D2 brane bound
state in a nontrivial configuration encoded in the $Y_a$ adjoints
of $U(k)$.

Moreover, this term is needed to describe the ordinary
contribution to the mass of the 2-0 strings when the D2 and D0
branes are separated in transverse directions. That is, there must
be a term in the low energy effective theory of the form
\begin{equation} \Tr \left( A_a Y_a^\dag Y_a A_a^\dag + Z_a^\dag
A_a A_a^\dag Z_a \right). \end{equation} We will soon see that the
F-term, $\left| \partial W \right|^2$, only gives an appropriate
mass to the bifundamentals $A_1$ when there is a distance $(Y_2)_i
- (Z_2)_j \neq 0$ between a pair of eigenvalues, however the $A_2$
mode must also receive a mass. Later, it will become clear the
condition $Z_a^\dag A_a = 0$ is redundant, and hence the
associated terms in the action can be absorbed into the already
existing ones, but the constraint \begin{equation} A_a Y_a^\dag =
0 \label{extra} \end{equation} is new, and must be imposed in
addition to the ordinary quiver conditions. This is similar to the
mass term needed for the 6-0 strings generated by separation of
the branes in the direction transverse to the Calabi-Yau. Note
that there is no off-diagonal residual gauge symmetry, since the
form of $B_2$ breaks it to exactly $U(N) \times U(k)$; this would
be an issue for $GL(N+k, \C)$, however.

The correct bundle over the classical moduli space can be found by
including the fields that are relevant for motion normal to the
Calabi-Yau, playing an analogous role to $\vp$ and $\phi$ in the
D0 theory. By the transversal $SO(7)$ rotational symmetry broken
from $SO(9,1)$ by the presence of the D2 brane, there must exist 3
additional low energy 0-2 modes, a complex chiral field,
$\widetilde{A}$, and a real vector, analogous to $\phi$. The
vector multiplet is associated to the unbroken off-diagonal
component of $GL(N+k, \C)$, and only its associated D-term
survives in the Kahler description, as we have seen. The presence
of $\widetilde{A}$ is important for finding the correct 1-loop
determinant at the fixed, although it is vanishing for all BPS
solutions.

The dynamical topological multiplets, describing the motion
(adjoint fields) and tachyons (bifundamentals) of the D0 branes
are
\begin{equation*}
\begin{split}
\delta Z_i = \psi_i , \ \delta \psi_i = [\phi, Z_i] \\
\delta \varphi = \zeta, \ \delta \zeta = [\phi, \varphi] \\
\delta q = \rho, \ \delta \rho =  \phi q \\
\delta B = \beta, \ \delta \beta =B \phi - \phi' B \\
\delta A_a = \alpha_a, \ \delta \alpha_a = A_a \phi' - \phi A_a \\
\delta \widetilde{A} = \widetilde{\alpha} \\ \delta
\widetilde{\alpha} = \widetilde{A} \phi' - \phi \widetilde{A}
\end{split}
\end{equation*} where $i = 1,2,3$ and $a=1,2$, and $B, A_a$ are the lowest lying 2-0 string modes, with masses
$m_B = - m_A <0$ at all values of the background Kahler moduli.
There are auxiliary multiplets \begin{equation*} \begin{split} \delta \phi = 0 \\
\delta \overline{\phi} = \eta \\ \delta \eta = [\phi,
\overline{\phi}], \end{split} \end{equation*}  as before. We
regard the D2 brane moduli as frozen due to the noncompactness of
their worldvolume, but they can still be derived from the T-dual
D4/D0 system as:
\begin{equation*} \begin{split}
\delta Y_a = \xi, \ \delta \xi = [\phi', Y_a] \\
\delta J = \upsilon, \ \delta \upsilon = \phi' J \\
\delta K = \kappa, \ \delta \kappa = K \phi' \\
\delta \phi' = 0, \end{split} \end{equation*} where the completion
to the full ten dimensional theory is ignored, as we are not
concerned with with anti-ghost bundle over this moduli space,
having already chosen a particular point due to the noncompactness
of the wrapped 2-cycles.

The topological nature of the partition function means that the
Euler character of the obstruction bundle is a deformation
invariant when it satisfies the proper convergence properties. In
particular, the cubic terms in the superpotential are sufficient
to determine the exact result, as higher order corrections, even
if they exist, will not affect the answer, although they may
correct the geometry of moduli space. The superpotential can be
read off from the quiver diagram, including the usual D0
Chern-Simons like term as well as the natural superpotential of
the D2/D0 system, giving
\begin{equation} W = \Tr \left( \Omega_{ijk} Z_i Z_j Z_k +
\epsilon_{ab} Z_a A_b B + \epsilon_{ab} A_a Y_b B + q K B \right),
\label{W026}
\end{equation} which implies the F-flatness conditions,
\begin{equation} \begin{split} [Z_1, Z_2] = 0, \ [Z_3, Z_a] = A_a B, \ B Z_a =
Y_a B, \\ Z_1 A_2 + A_1 Y_2 = Z_2 A_1 + A_2 Y_1 + q K , \ K B = 0
, \ B q = 0. \end{split} \end{equation}

The moment maps are \begin{equation} \begin{split} \sum [Z_i,
Z_i^\dag] + \sum A_a A_a^\dag - B^\dag B + q q^\dag = r I_N, \\
\sum [Y_a, Y_a^\dag] - \sum A_a^\dag A_a + B B^\dag + K^\dag K - J
J^\dag = r' I_M, \end{split} \end{equation} which square to give
the D-term contribution to the topological matrix model. Note in
particular that $r' > r$ so that the usual 2-0 string is
classically tachyonic, as it must be.

The collective coordinates, $Y_a$, of the D2 brane will encode
their configuration as bound flux in the D6, which we will regard
as a fixed asymptotic condition, due to the noncompactness of the
D2 worldvolume. There is also the F-term constraint for the D6/D2
system,
\begin{equation} [Y_1, Y_2] = J K. \end{equation}

First let us understand why these are physically the correct
conditions. The FI parameters serve to give masses to the
bifundamentals via terms of the form $- 2 r \Tr A_a A_a^\dag$, and
we find that the 0-2 tachyon has bare mass $-2 (r'-r) <0$ for all
values of the background B-field. There is a pair of massive
fields, $A_a$, in the 0-2 spectrum with mass $2(r'-r)$ that allow
us to write the usual superpotential. The 6-2 string $K$, which is
also tachyonic with mass $-2r' < -2 (r'-r)$, will be the first to
condense, dissolving the D2 branes into flux on the D6.

Recall that the D6/D2 system is T-dual to D4/D0, and the bound D0
flux is described by the ADHM construction \cite{ADHM}. Therefore
the combination $\sum [Y_a, Y_a^\dag] + K^\dag K - J J^\dag$ is an
$M \times M$ matrix with positive eigenvalues, which results in a
large positive contribution, from the term $\Tr B^\dag (\sum [Y_a,
Y_a^\dag] + K^\dag K - J J^\dag) B$ in the action, to the
effective mass of $B$ in the vacuum with nonzero $K$. Therefore we
find that $B=0$ in the supersymmetric vacuum, which also agrees
with the F-term condition. This can be thought of as a quantum
corrected mass for $B$ after integrating out the heavy field, $K$.

As can easily be seen from the structure of the D-terms, the
effective masses are $m_{A} = -m_B$, hence these bifundamentals
now condense, binding the D0 branes to the D2. Putting everything
together, we see that the background D2 configuration is described
by a Young diagram, $\lambda$, with $M$ boxes, encoded in the
matrices $Y_a$, as expected. The D0 collective coordinates obey
$[Z_i,Z_j] = 0$, giving a difference, $\eta$, of three dimensional
partitions, and the D-term can only be satisfied when the interior
corners are exactly the image of the map $A_a$ from $\C^M
\rightarrow \C^N$.

Moreover, there is a holomorphic equation which implies that
\begin{equation}\label{mesh} Z_1 A_2 + A_1 Y_2 = Z_2 A_1 + A_2 Y_1 .\end{equation} It is also
clear from the discussion of the effective mass of the
bifundamentals that $Y_a A_a^\dag=0$ for $a=1, \ 2$. This means
that only one of the terms on each side of \eqref{mesh} can be
nonvanishing. The geometric interpretation in terms of crystal
configurations is that the $z_3=0$ plane of $\eta$ together with
$\lambda$ forms a Young diagram, and by the commutativity of the
$Z_i$, this implies that $\eta$ is exactly a three dimensional
partition in the background of $\lambda \times z_3$ axis, as
expected from \cite{topcrystal}!

To further check that \eqref{mesh} is the correct F-flatness
condition even for non-torus invariant points in the BPS moduli
space, note that shifting the positions of the D0 and D2 brane
together by sending $Z_i \mapsto Z_i + x_i I_N$ and $Y_a \mapsto
Y_a + x_a I_k$ doesn't change the solutions for the $A_a$, as
expected. It is easy to confirm that in the generic branch of
moduli space, specifying $3 N$ independent eigenvalues of mutually
diagonalizable $Z_i$ and likewise $2 k$ distinct eigenvalues of
$Y_a$ totally determines the $A_a$ up to gauge equivalence. This
is because \eqref{mesh} essentially allows one to solve for $A_2$
in terms of $A_1$, and the condition \eqref{extra} together with
the D-terms fix $A_1$ up to the $U(1)^{N+k-1}$ symmetry left
unbroken by the choice of $Z_i$ and $Y_a$.

The path integral can be calculated as before by introducing a
toric regulator. The fields transform as before under the $U(1)^3$
action on $\C^3$, with \begin{equation} A_a \To e^{i \epsilon_a }
A_a, \ \ \ \ B \To e^{i \epsilon_3} B, \ \ \ \ \widetilde{A} \To
e^{i \sum \epsilon_i} \widetilde{A}, \end{equation} and
consistently for the anti-ghosts. Therefore the full result for
the determinant is given by \begin{equation*}
\begin{split} \frac{(\ad \phi + \epsilon_1 + \epsilon_2)(\ad \phi +
\epsilon_1 + \epsilon_3)(\ad \phi + \epsilon_2 + \epsilon_3) (\phi
- \phi' + \epsilon_1 + \epsilon_2) (\phi' - \phi + \epsilon_3+
\epsilon_1)}{(\ad \phi + \epsilon_3)(\ad \phi + \epsilon_2)(\ad
\phi + \epsilon_1) (\phi - \phi' + \epsilon_3) (\phi' - \phi +
\epsilon_2)} \\ \times \frac{(\phi' - \phi + \epsilon_3+
\epsilon_2)}{(\phi' - \phi + \epsilon_1)} \frac{(\phi - \epsilon_1
- \epsilon_2-\epsilon_3)(\phi - \phi')}{(\phi)(\phi-\phi' +
\epsilon_1 + \epsilon_2+\epsilon_3)(ad \phi + \epsilon_1 +
\epsilon_2+\epsilon_3)},
\end{split}\end{equation*} which precisely cancel when the toric
weights sum to zero, and we include the Vandermonde from the
$\phi$ integral. The measures from the fields governing motion of
noncompact objects have not been included since they are frozen,
not integrated.
\end{subsection}

\begin{subsection}{Multiple asymptotics and a puzzle}

Now we would like to understand the quiver matrix model that
describes the bound states of D0 branes to our D6 when D2 charge
is turned on in more than one of the three toric 2-cycles. The
index of BPS states on this theory, with a fixed configuration of
the nondynamical fields associated to motion of noncompact
objects, should give exactly the topological vertex of
\cite{AKMV}, \cite{qfoam}. Generalizing the quiver, it is again
clear that the frozen D2 adjoint fields should give a stable
representation of the constraint $[Y_1, Y_2]=0$; these are the
asymptotic Young diagrams appearing in the topological vertex,
$C_{\mu \nu \eta}$. See Figure ~\ref{vq} for the complete quiver
diagram including all fields.

The quiver is given as before, with a node, $U(k_i)$, for each of
the three stacks of frozen D2 branes, and bifundamentals 2-2'
strings, which have the same low energy field content as a T-dual
pair of 4-0 bifundamentals. Only the cubic terms in the
superpotential are relevant, and the possible terms allowed by the
quiver (which also respect the obvious rotational symmetries in
$\C^3$) are
\begin{equation} W_0 = \epsilon_{i j k} \Tr \left( Z_i Z_j Z_k
\right) + \epsilon_{i j k} \Tr \left( B^i \left( Z_j A^i_k + A^i_j
Y_k \right) \right) + \epsilon_{i j k} \Tr \left(J^i_j B^i A^j_k
\right) + K_i B_i q, \end{equation} for the D0 degrees of freedom
and
\begin{equation} W_2 = \Tr \left( J^1_2 J^2_3 J^3_1 - J^1_3 J^3_2 J^2_1
\right) + \epsilon_{i j k} \Tr \left( J^i_j Y^i_k J^j_i \right) +
K_i J^j_i \widetilde{K}_j,
\end{equation} where we have only included couplings of the
dynamical fields. The frozen 6-2 modes must satisfy the equations
of motion obtained from $\partial W_{frozen} = 0$, where
\begin{equation} W_{frozen} =  K_i \widetilde{Y}^i \widetilde{K}_i + \epsilon_{i
j k} \Tr \left(\widetilde{Y}^i Y^i_j Y^i_k \right) ,
\end{equation} where the $\widetilde{Y}^i$ are exactly 0 in the
vacuum, being associated to motion transverse to the D6 brane, and
have been introduced simply to be able to write the
superpotential.  Note that it is possible to rescale the fields,
while preserving rotational invariance in $\C^3$, in such a way to
set all of the relative coefficients in the superpotential to
unity.

The physical moduli space is the Kahler quotient of the resulting
algebraic variety by $U(N) \times U(k_1) \times U(k_2) \times
U(k_3) \times U(1)$, where the overall $U(1)$ acts trivially on
all of the fields. This means that there is a D-term in the action
given by the square of the equations \begin{equation}
\begin{split} \sum_{i=1}^3 [Z_i, Z_i^\dag] + \sum_{i \neq a = 1}^3
\left( A^i_a \right) \left( A^i_a \right)^\dag + q q^\dag = r_N I_N \\
\sum_{a \neq i} \left( [Y^i_a, {Y^i_a}^\dag] - {A^i_a}^\dag A^i_a
+ J^a_i {J^a_i}^\dag - {J^i_a}^\dag J^i_a \right) + I^i {I^i}^\dag
- {\widetilde{I}}^{i \dag} \widetilde{I}^i = r_i I_{k_i} \\
q^\dag q + \sum_{i=1}^3 {I^i}^\dag I^i - \widetilde{I}^i
\widetilde{I}^{i \dag} = r_6, \label{fullDterm}
\end{split}
\end{equation} where the FI parameters, $r$, are determined by the
background values of the Kahler moduli in a complicated way. We
will later graph the loci where various combinations of these
Kahler parameters of the moduli space vanish, which can be found
by looking for walls of marginal stability where the central
charges of some of the constituent branes align.

The fact that we are working in a local geometry obtained as a
noncompact limit far from the Gepner point of the global
Calabi-Yau again means that there will exist residual off-diagonal
D-terms. Collecting all of the relevant equations, one has that
\begin{equation} \begin{split} A^i_j {Y^i_j}^\dag + Z_j^\dag A^i_j
+ A^j_i J^{j \dag}_i = 0 \\ J^i_j Y^{i \dag}_k + Y^{j \dag}_i
J^i_k = 0 \label{odiag}
\end{split}
\end{equation}

The 2-2' bifundamental strings are localized in $\C^3$, stretched
along the minimum distance between the orthogonal 2-branes (and
thus living at any non-generic intersection), so they are also
dynamical fields. Looking at the form of $W_2$, it is possible to
see that the 2-6 system without D0 branes has no unlifted moduli
after fixing the frozen degrees of freedom, that is, the
semi-classical BPS moduli space is a point. This can be easily
seen, since if the D2 branes are separated by a large distance in
$\C^3$, all of the $J^i_j$ become heavy. Thus the moduli space we
are interested in is indeed the effective geometry as seen purely
by the 0-brane probes.

There is a new feature in this quiver, since it is not immediately
obvious which of the 6-2 and 2-2' strings should given background
VEVs. We will find that depending on the values of the background
Kahler moduli and B-field, the BPS ground states will live in
different branches of the moduli space of these fields, which are
frozen from the point of view of local D0 dynamics. The resulting
effective quivers for the D0 degrees of freedom will be different,
although the final computation of the Euler character turns out to
be independent. In the next section, it will become clear what
this means in terms of quantum foam. The alternative quiver
realizations will exactly correspond to the resolutions, related
by flops, of the blown up geometry experienced by the D0 branes,
viewed as probes. It is thus very natural that the effective
Kahler structure depends on the background Kahler moduli, via the
Fayet-Iliopoulos parameters.

To get a feel for the way this works, consider the simplest
example with two nontrivial asymptotics, $C_{\Box \Box \cdot}$.
There are two equivariant configurations of the D6-D2 system that
are consistent with the F-term condition, $\partial W_{frozen}=0$,
namely $Y^i_j = 0$ and either $J^1_2 = 0, J^2_1 = 1$ or $J^2_1=0,
J^1_2 = 1$. Only one of these solutions can satisfy the D-term
constraint for the D2 $U(1)$ gauge groups, for any given value of
the FI parameters. For definiteness, let us focus on the latter
solution. Then $N$ D0 branes probing this background will be
described by the quiver as before, with the new condition that
\begin{equation} Z_3 A^1_2 + A^2_3 J^1_2 = Z_2 A^1_3, \label{tra}
\end{equation} whose equivariant fixed points can be interpreted
as a skew three dimensional partition generated by the commuting
$Z_i$ in the usual way. The vectors $A^1_3$ and $A^2_1$ form the
interior corners of the partition, saturating the moment map
condition, while $A^1_2 = 0$ by \eqref{offdiag}, and $A^2_3$,
interpreted as a vector in $\C^N$, is simply given be $Z_2 A^1_3$
by equation \eqref{tra}. Using the further holomorphic equation
that
\begin{equation} Z_1 Z_2 A^1_3 = Z_1 A^2_3 J^1_2 = Z_3 A^2_1
J^1_2, \end{equation} we see that this means the skew partition
exactly fits into the empty room crystal configuration associated
to $C_{\Box \Box \cdot}$.

The essential idea extends to arbitrary background D2 brane
configurations. In particular, there are usually many choices in
determining the frozen fields, which depend of the FI parameters,
and thus on the background Kahler moduli. One way to solve the
constraints is to start with one asymptotic, and condense the
tachyonic 2-6 field, being sure to tune the FI parameters so that
it is along the direction of steepest descent in the potential.
This procedure will give new contributions to the effective masses
of the 2-2' strings, and there will again be some region in the
background moduli space where they become unstable and condense.
Finally, some of the 2-2'' and 2'-2'' fields condense. The D-term
condition for each first $U(k_1)$ is supported by the 2-6
fundamental, while the D-terms of the two successive $U(k_i)$ are
saturated by the condensed 2-2 bifundamental modes.

The above rather complicated sounding procedure, when unraveled,
simply acts as a mechanism to obtain the correct equation relating
the vectors of $U(N)$ corresponding to the edge of the background
D2 branes in the crystal, starting with only a cubic
superpotential (and thus quadratic equations $\partial W=0$). If
we regard the dynamical 2-0 bifundamentals only as vectors of
$U(N)$, then relations are schematically of the form $Z_1^a Z_2^b
Z_3^c A = Z_1^d Z_2^e Z_3^f A'$, where the powers of $Z_i$ are
determined by the shape of the intersecting asymptotic Young
diagrams. The various quivers for a given vertex, $C_{\mu \nu
\eta}$, are associated to different configurations of the frozen
fields which lead to the same final conditions of this form.
Physically speaking, the effective theory for the $Z_i$, obtained
by integrating out all the bifundamental matter, is independent of
the attractor tree, as long as no walls of marginal stability are
crossed. However the full action with cubic superpotential that
"completes" the higher order effective action can differ as we
vary the background moduli.

A particularly simple example is $C_{\Box \Box \Box}$, which turns
out to have four quiver representations, one of which is $A^i_j
\propto A^j_i$, as $\C^N$ vectors, which obey the equations $Z_k
A^i_j = Z_j A^i_k$. We will soon see how all of these facts come
together in a compelling physical picture, which gives a robust
realization of the quantum foam geometries of \cite{qfoam} as the
physical geometry experienced by probe D0 branes.

\end{subsection}

\end{section}

\begin{section}{Effective geometry, blowups, and marginal stability}

%\textbf{Aside:}

%Is it possible to determine the Kahler modulus in terms of the
%background moduli? Ie. quantitatively, rather than simply which
%regime separated by the flop? Can we see this from the metric on
%the D0 moduli space?

%Consider the near horizon geometry of D6-D2 configuration. Imagine
%there is a probe BPS D0 brane in the near horizon regime. It must
%see a blow up of the CY as effective geometry!

%\textbf{End aside}

Consider the quiver describing a single D0 brane in the vertex
with one asymptotic 2-brane bound state encoded in the Young
diagram, $|\mu| = k$. Then the relevant dynamical fields are the
D0 coordinates and the 2-0 tachyons, which satisfy the F-flatness
relation
\begin{equation} A_1 \left(Y_2-z_2 \right) = A_2
(Y_1-z_1),\end{equation} where $Y_a(\mu)$ are fixed matrices. It
is easy to check that the projectivization of this holomorphic
moduli space obtained by moding out by the residual nontrivial
$\C^\times$ gauge symmetry left unbroken by the $Y_a$ is exactly
the blow up of $\C^3$ along the ideal defined by $\mu$. Suppose
that all of the parallel 2-branes are separated in the transverse
$z_1$-$z_2$ plane. Then, working in the holomorphic language, we
can use $GL(k, \C)$ to simultaneously diagonalize the $Y_a$, and
in that basis we therefore find \begin{equation*} (A_1)_m \left(
(y_2)_m - z_2 \right) = (A_2)_m \left( (y_1)_m - z_1 \right), \ \
\ \ m = 1, ..., k
\end{equation*} These $k$ equations define the blow up of $\C^2$ at
$k$ distinct points; the $z_3$ direction is a simple product with
the resulting smooth, non Calabi-Yau 4-manifold. As we imagine
bringing the D2 branes on top of each other, exploring other
branch of the Hilbert scheme of $k$ points in $\C^2$, the
effective geometry experiencing by the 0-brane will be the blow up
along that more complicated ideal sheaf. The smoothness of the
Hilbert scheme in two complex dimensions (which is the moduli
space of the 2-branes bound to a D6) assures one that no
subtleties can emerge in this procedure, as can be confirmed in
specific examples.

More generally, using $N$ 0-branes as a probe, we would see a
geometry related to, but more complicated than, the Hilbert scheme
of $N$ points in this blow up variety. In particular, the 0-branes
are affected by residual flux associated to the blown up 2-branes,
so that globally they live in sections of the canonical line
bundle associated to the exceptional divisors, rather than the
trivial bundle. More abstractly, the associated sheaves fit into
an exact sequence
\begin{equation*} 0 \To {\cal I} \To {\cal L} \To i_*{\cal O}_Z \To
0, \end{equation*} where ${\cal L}$ is the nontrivial line bundle,
and $Z$ encodes the bound point-like subschemes. This doesn't
change the nature of the Euler character of the moduli, which is
determined by the local singularities. Intriguingly, there is a
further refinement which naturally associates a size of order
$g_s$ to the exceptional divisors. It is impossible for more than
a small number (typically two in our examples) of D0 branes to
``fit'' on the exceptional divisor. It would be interested to see
if this feature also emerged in the study of the exact BPS moduli
space of $N$ 0-branes on a Calabi-Yau in the small volume limit.

In many examples with two or three nontrivial asymptotics, there
exist multiple resolutions of the blown up geometry, as shown in
the simple example of $C_{\square \square \cdot}$ (see Figure
~\ref{blowup}). We can see this as well in trying to write down
the quiver description of these moduli spaces: there are different
choices of background D2 moduli which, however, all have equal
Euler characteristic. In fact, there is a clear physical reason
for this, to be found in a more detailed analysis of the lines of
marginal stability. Depending on the values of the background
Kahler moduli, different bifundamental tachyons will condense
first. The index of BPS states is independent of which attractor
tree \cite{DM} the system follows, as long as no lines of marginal
stability are crossed.

\begin{figure}
\begin{center}
$\begin{array}{c@{\hspace{1in}}c}
 \epsfig{file=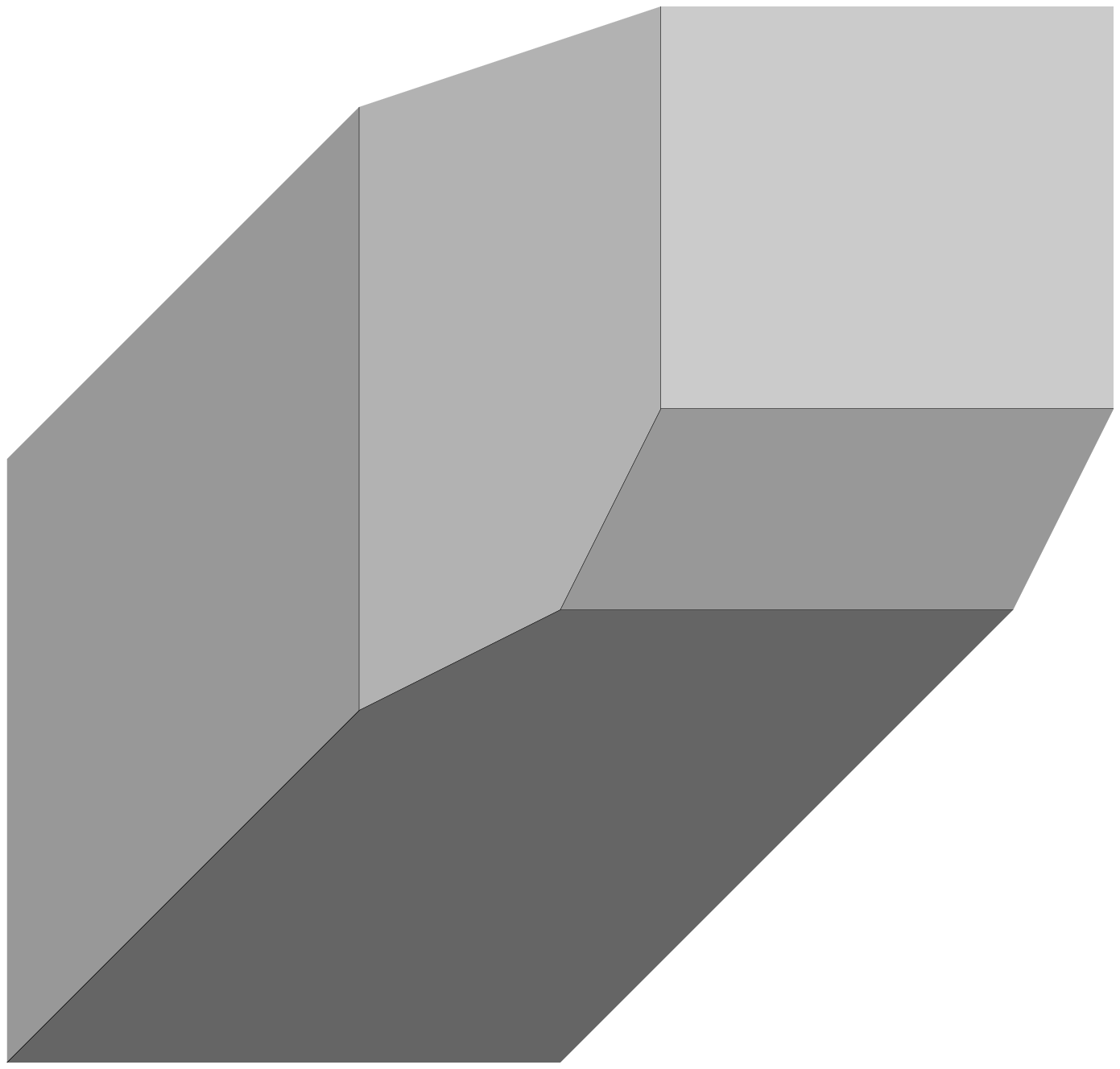,width=5cm} &   \epsfig{file=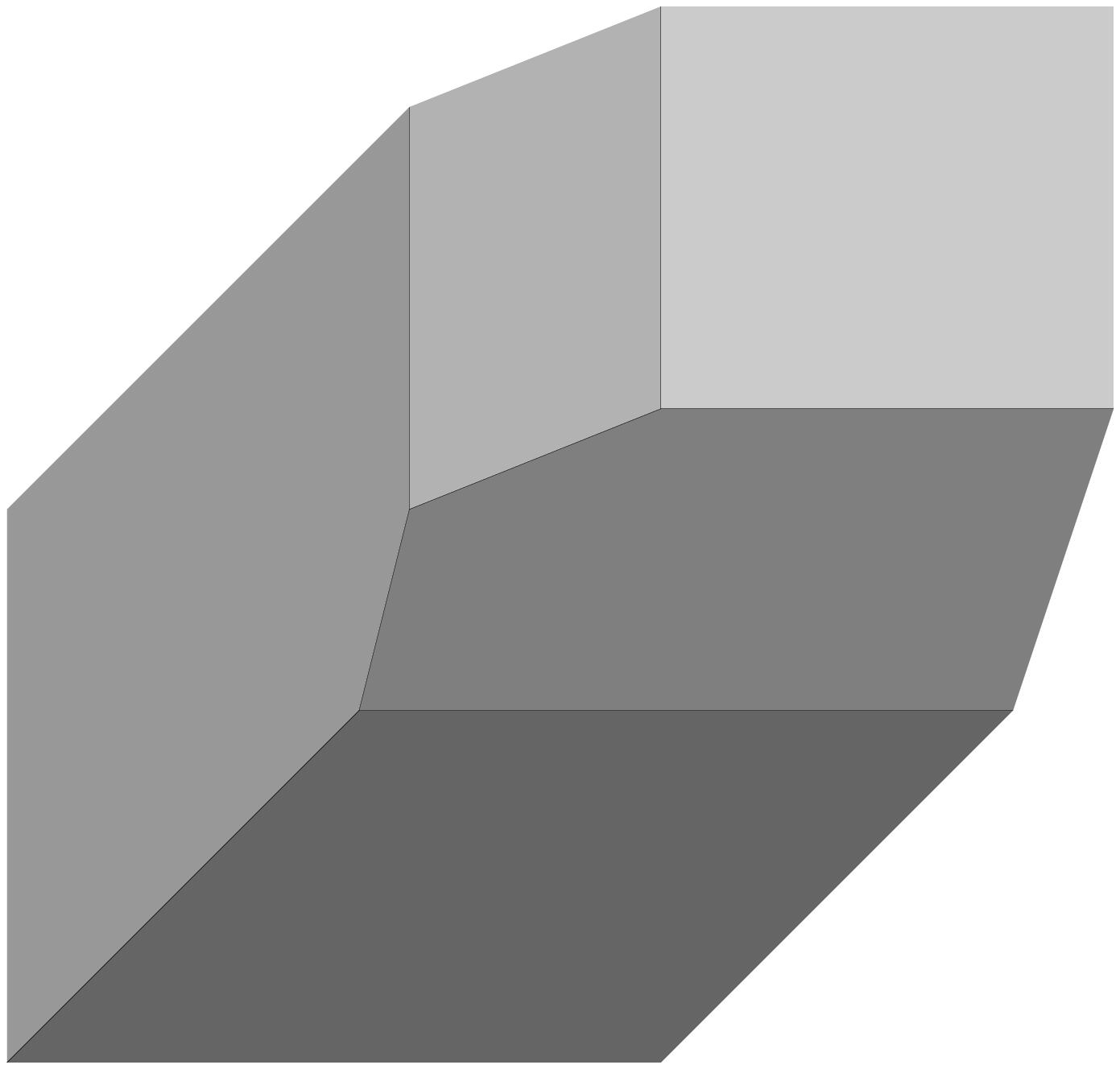,width=5cm}
 \end{array}$ \caption{The two blowups associated to $C_{\Box \Box \cdot}$,
related by a flop.} \label{blowup}
\end{center}
\end{figure}

The partition function for the general vertex reduces to the Euler
characteristic of the anti-ghost bundle over the classical moduli,
because of the topological nature of the matrix integral. As we
saw in the previous section, this Euler character can be localized
to the fixed points of a $T^3$ action induced on the moduli space,
and the weights at the fixed points are all equal when the torus
twist preserves the $SU(2)$ holonomy, as shown by computing the
1-loop determinant near the fixed point.

We now investigate the properties of the moduli space itself,
which it the Kahler quotient of a holomorphic variety defined by
quadratic equations, resulting from the cubic superpotential. The
three F-term equations,
\begin{equation} Z_i A^k_j + A^k_i Y^k_j + A^j_i J^k_j = Z_j A^k_i
+ A^k_j Y^k_i + A^i_j J^k_i, \label{fullhol}
\end{equation} where $i,j,k$ are distinct and {\it not} summed
over, together with $[Z_i, Z_j]=0$ serve to define the holomorphic
moduli space of the dynamical low energy degrees of freedom.

In our noncompact setup, we have not specified the boundary
conditions for the 2-branes, which are frozen in the local
analysis. If they are free to move in the transverse directions,
generically they will not intersect, and the effective geometry
probed by a BPS 0-brane will be the blowup along the wrapped
2-cycles, as can be seen by going to a complex basis in which
$Z_i$ and $A^i_j$ are all diagonal. The 2-2' strings will become
massive, and exit the low energy spectrum, when the branes are
separated, and \eqref{fullhol} becomes $k_1 + k_2 + k_3$ equations
describing the local geometry $\C \times {\cal O}(-1) \To \P^1$
near each of the blown up 2-cycles.

It is also very interesting to consider compactifications in which
the D2 branes wrap rigid or obstructed 2-cycles. In that case,
they will have a nontrivial intersection, and a more sophisticated
analysis is needed. As we will show in the following examples, the
effective geometry is again the blow up along the wrapped
subschemes. It is here that multiple smooth resolutions of the
geometry may exist, corresponding to different attractor trees in
which the order in which the tachyonic fields condense changes. No
walls of marginal stability are crossed, hence the index of BPS
remains constant.

The supersymmetric bound states we are discussing only exist when
the asymptotic B-field is sufficiently large. From the perspective
of Minkowski space, as the B-field flow down to the attractor
value, various walls of marginal stability will be crossed. Thus
the full geometry is multi-centered. Certain features of the
attractor flow tree are visible in the quiver matrix model, so we
first examine them from supergravity.

The central charge of a D6/D4/D2/D0 bound state described by a
complex of sheaves ${\cal E}^*$ over the Calabi-Yau is, at large
radius,
\begin{equation} \label{cc} Z({\cal E}^*) = \int e^{-\omega} {\it ch} ({\cal E}^*) \sqrt{Td(X)}, \end{equation}
where $\omega = B + i J$ is the Kahler $(1,1)$-form on $X$
\cite{Aspinwall}. At small volumes, this would receive worldsheet
instanton corrections, hence the following analysis should only be
expected to indicate the qualitative behavior of the attractor
flow. This is sufficient for our purposes, since the local limit
automatically assumes large volume.

In our case of $k_i$ 2-branes and $N$ 0-branes bound to a D6 brane
in the equivariant vertex, the central charge \eqref{cc} is
computed as
\begin{equation} \int e^{\sum t_i \omega_i} \left( 1 + k_1
\omega_2 \wedge \omega_3 + k_2 \omega_1 \wedge \omega_3 + k_3
\omega_1 \wedge \omega_2 + N \omega_1 \wedge \omega_2 \wedge
\omega_3 \right) = t_1 t_2 t_3 + \sum t_i k_i + N,
\end{equation} where the $t_i$ are Kahler parameters associated to
toric legs. They may be independent or obey some relations,
depending on the embedding of the vertex into a compact 3-fold.

From this one can check the well known fact that the 6-0 BPS bound
state only exists when $\arg (t^3) > 0$, that is $|B_{z \bar z}| >
\left( 1/\sqrt{3} \right) |g_{z \bar z}|$. Similar calculations
have been done in \cite{DM} when there are also D2 branes. We are
most interested in the behavior of the 2-branes as the moduli
approach the attractor values. Note that in a local analysis, the
attractor flow itself cannot be determined without knowing the
compactification; this is obvious because when the $t_i$ become
small, the vertex is no longer a good approximation.

 All of these
details of the global geometry are hidden in the FI parameters and
the VEVs of the frozen fields appearing in the effective action
for the 0-brane degrees of freedom. This is consistent with the
expectation that the theory is defined at the asymptotic values of
the moduli, where the 2-brane fields are heavy and their
fluctuations about the frozen values can be integrated out. The
topological nature of the matrix models guarantees that the only
effect will be a ratio 1-loop determinants, which in fact cancel
up to a sign.

As the moduli flow according to the attractor equations from the
large B-field asymptotic limit needed to support the 6-brane BPS
states, there are various possible decays. To find the walls along
which they occur, one finds the condition for the central charges
of the potential fragments to align. We find that in our examples,
the 0-branes first split off.

The remaining D6/D2 bound state will itself decay at another wall
of marginal stability. Depending on the asymptotic value of the
moduli, the 2-branes wrapped one of the three toric legs will
fragment off. Because the relevant Kahler moduli space is six real
dimensional even in the local context, the full stability diagram
is difficult to depict on paper, so we will content ourselves with
the representative cross-section shown in Figure ~\ref{stable}.
The charges in this example are denoted by \begin{equation*}
\begin{split} \Gamma = \omega_1 \wedge \omega_2 \wedge \omega_3 +
\omega_1 + \omega_2 + 1 \\ \Gamma_{6/2} = \omega_1 \wedge \omega_2
\wedge \omega_3 + \omega_1 + \omega_2 \\ \Gamma_1 = \omega_1 \\
\Gamma_2 = \omega_2. \end{split} \end{equation*} We choose the
background Kahler moduli to be $$t_1 = 16 i + x, \ \ t_2 = 14 i +
y, \ \ t_3 = 4 i + 4, $$ and plot the $x, \ y$ plane which
parameterizes the asymptotic value of the B-field along the curves
wrapped by the D2 branes.

\begin{figure}
\begin{center}
\epsfig{file=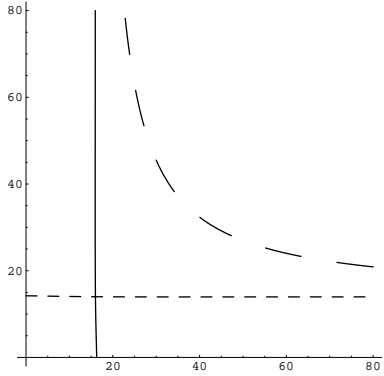,width=5cm} \caption{Walls of marginal
stability for the decays $\Gamma \To \Gamma_0 + \Gamma_{6/2}$
(dashed line), $\Gamma_{6/2} \To \Gamma_1 + \Gamma'$ (dotted), and
$\Gamma_{6/2} \To \Gamma_2 + \Gamma''$ (solid).} \label{stable}
\end{center}
\end{figure}

It is immediately apparent that as the B-field in the $z_1$ and
$z_2$ planes decrease along the attractor flow, the 0-branes
separate along the first curve shown, and then the D6/D2 fragment
intersects one of the two 2-brane curves. We shall see below how
this phenomenon is imprinted on the matrix model. Therefore, in
our local analysis, the 2-brane degrees of freedom are frozen in a
pattern which depends on the attractor flow tree. It is natural
that the indices calculated for the different effective 0-brane
geometries are equal, since we see that they arise from the same
theory in the global Calabi-Yau geometry.

\begin{subsection}{Probing the effective geometry of a single 2-brane with $N$ 0-branes}

Let us first understand a very simple example from this point of
view, namely the effective geometry induced by a single D2 brane,
corresponding to the vertex $C_{\Box \cdot \cdot}$. Consider
probing the induced geometry wrapped by the D6 brane using $N$
0-branes. The F-term equation \eqref{mesh} tells us that $Z_1 A_2
= Z_2 A_1$, where we set the D2 position to 0 without loss of
generality. Ignoring the $z_3$ direction for now, as it simply
goes along for the ride, we see that this can be related to the
quiver for a D4/D2/D0 bound state in ${\cal O}(-1) \To \P^1$,
which lifts to D6/D4/D0 in the three dimensional blow up. The D4
brane is equivalent to the line bundle mentioned before, since it
can be dissolved into smooth flux on the 6-brane worldvolume.

To see this, consider the two node quiver for ${\cal O}(-1) \To
\P^1$ obtained by dropping one dimension from the quiver of
resolved conifold. Setting the field $D = \left( I 0 \right)$ for
quiver charges $N$ and $N+1$ splits $U(N+1) \To U(N) \times U(1)$,
and allows one to define \begin{equation} Z_a = C_a D,
\end{equation} and $A_a$ to the be the $U(1)$ piece of $C_a$. Then
the relation $C_1 B C_2 = C_2 B C_1$ implies that $Z_1 A_2 = Z_2
A_1$ and $[Z_1, Z_2]=0$, as desired.

The two moduli spaces are not quite identical, however, because it
is not always possible to choose a gauge such that $D = \left( I \
0 \right)$, and conversely, given such a form for $D$, it is
impossible to satisfy the D-term condition when $Z_a = 0$.
Therefore the structure of the moduli spaces near the zero section
of ${\cal O}(-1)$ differ. In particular, it is easy to see that if
$Z_1=Z_2=0$ in some $M$ dimensional subspace of $\C^N$ then there
is a $\P^1$ of $A_1, A_2$ if $M=1$, a single point if $M=2$, and a
violation of the D-term if $M>2$. This is the sense in which the
resolution is small - only a single 0-brane can fit conformably.

Lifting to the full $\C^3$ adds one complex dimension to the
2-brane, hence the even far from the exceptional divisor the
0-branes will see a $U(1)$ flux turned on in the D6 gauge theory.
This has no effect on the moduli space of a single BPS D0, but
will twist the global structure of the $N$ D0 moduli space as
follows.

Ordinarily, the Hilbert scheme of points is a kind of non-smooth
``resolution'' of the classical space of $N$ points in a 3-fold,
$\Sym^N (X)$. For simplicity, consider the case of $\Sym^2 (X)$,
which looks like the bundle ${\cal T}/\Z_2 \To \Delta$ near the
diagonal embedding, $\Delta: X \hookrightarrow \Sym^2 X$, where
${\cal T}$ is the tangent bundle, with the natural $\Z_2$ action
$z_i \To - z_i$ in local coordinates on the fiber. The $z_i$
should be thought of as the small relative separation of the two
points. In this special case, the Hilbert scheme
$\operatorname{Hilb}^2 (X)$, is locally the smooth resolution of
the $\Z_2$ orbifold, which replaces the fiber by ${\cal O}(-2) \To
\P^2$.

Therefore near the diagonal, the Hilbert scheme of two points on
$X$ is given by \begin{equation*} \begin{array}{ccc} {\cal O}(-2)
\otimes {\cal K} &
\longrightarrow & \P {\cal T} \vspace{.5 mm} \\ & & \downarrow \vspace{0.5mm} \\
& & X,
\end{array}
\end{equation*} where the ${\cal O}(-2)$ bundle is fibered
trivially over the base, and ${\cal K}$ is the canonical bundle
over $X$. When a $U(1)$ flux on the D6 worldvolume is turned on,
carrying D4 charge, the ${\cal O}(-2)$ of the Hilbert scheme
becomes fibered over the diagonal in the sense of ${\cal L}$, the
$U(1)$ bundle over $X$. That is, we should write ${\cal L}_X
\otimes {\cal K}_X \otimes {\cal O}(-2)_{\P^2}$ in the above
diagram of the geometry where the points are close together. This
is compactified in a natural way to the full Hilbert scheme with a
background line bundle, and a similar situation holds for $N > 2$
D0 branes.

Intriguingly, the effective canonical class of $X$, in the sense
of the what line bundle appears in the above description of the
Hilbert scheme, turns out to be the trivial bundle in our case,
since ${\cal L}$ exactly cancels the fibration of ${\cal K}$ by
the nature of the blow up construction. It would be interesting to
understand  if there is any more robust way in which the blow ups
are effectively Calabi-Yau.

\end{subsection}

\begin{subsection}{Example 2: the geometry of $C_{\Box \Box
\cdot}$}

Let us work out the example of $C_{\Box \Box \cdot}$ in detail, as
it already possesses most of the new features. The classical
moduli of the D2 branes are the coordinates $y^1_2, y^1_3$ and
$y^2_1, y^2_3$ respectively. Depending on the details of the
compactification, which determine the superpotential for these
fields, it may be possible to separate the branes in the $z_3$
direction. In that case, the 2-2' strings develop a mass and exit
the low energy spectrum. Hence $N$ probe D0 branes will have a BPS
moduli space given by the $U(1)^2 \times SU(N)$ Kahler quotient of
\begin{equation}
\begin{split} \left(
Z_2 - y^1_2 \right) A^1_3 &= \left(Z_3 - y^1_3 \right) A^1_2 \\
\left(
Z_1 - y^2_1 \right) A^2_3 &= \left(Z_3 - y^2_3 \right) A^2_1 \\
[Z_i, Z_j] &= 0 . \end{split} \end{equation} For $y^1_3 \neq
y^2_3$, this looks locally, near each of the D2 branes, like the
geometry of $N$ D0 branes in the ``small'' blowup discussed
before. In particular, a single D0 probe sees exactly the blow up
geometry along the two disjoint curves.

More interesting  behavior occurs when the 2-branes cannot be
separated in this way, for example in the closed vertex geometry
in which the wrapped $\P^1$'s are rigid. This is more closely
connected with the equivariant result, which requires the wrapped
2-cycles to have such a non-generic intersection by torus
invariance. Here we must use the fact that \begin{equation} J^i_j
\left( A^i_j \right)^\dag = 0, \end{equation} which results from
the term \eqref{odiag} in the effective action.

The effective geometry of the $C_{\Box \Box \cdot}$ vertex, as
probed by an individual BPS D0 brane is therefore determined by
the
Kahler quotient of the variety ,\begin{equation} \begin{split} z_2 a^1_3 = a^2_3 j^1_2 \\
z_1 a^2_3 = z_3 a^2_1, \end{split} \end{equation} by $U(1) \times
U(1)$ under which the fields have charges \begin{equation*}
\begin{array}{ccccc}
z_i & a^1_3 & j^1_2 & a^2_1 & a^2_3 \vspace{0.1in} \\
0&1&1&0&0\\
0&1&-1&1&1 \end{array} \end{equation*} The first equation implies
that $(z_2, a^2_3)$ lives in the ${\cal O}(-1)$ bundle over the
$\P^1$ with homogeneous coordinates $(a^1_3: j^1_2)$. Moreover the
second equation fixes $(z_1, z_3)$ to be fibered in ${\cal O}(-1)$
on the compactification of the $a^2_3$ direction. This is shown in
the toric diagram (see Figure ~\ref{blowup}), which is exactly the
resolution of the blow up of $\C^3$ along the non-generically
intersecting $z_1$ and $z_2$ lines!

Reversing the roles of the two D2 branes will result in a flopped
geometry (which in this case is coincidentally geometrically
equivalent to the first). It seems reasonable to conjecture that
as one dials the background moduli, and thus implicitly the FI
parameters, such that the roles of the two branes are exchanged,
the attractor tree pattern changes without crossing a wall of
marginal stability. This behavior was observed in a related system
in \cite{DM}. We are using the intuitive identification between
the attractor flow and the line of quickest descent in the
potential of the matrix model.

\end{subsection}

\begin{subsection}{An example with three 2-branes}

The blowup of the vertex geometry along all three torus invariant
legs has a total of four smooth resolutions, related by conifold
flops. This is the right description when the legs are expected to
become rigid spheres in the global geometry, so they cannot be
holomorphically deformed away from the triple intersection. Let us
see explicitly how one of these blowups arises from the quiver
description of $C_{\Box \Box \Box}$.

Suppose that the 2-brane wrapping the $z_1$ plane has melted first
into the background D6, that is, $I^1 \neq 0$. Without any 0-brane
probes, the dynamical fields are the $J^i_j$, which satisfy F-term
conditions derived before requiring that $J^i_j J^j_i = 0$, $J^i_j
J^k_i = 0$, and $J^i_j I^i = 0$ for all distinct $i,j,k$. It is
easy to satisfy the D-term constrain, for appropriate values of
the FI parameters, by taking $J^2_1, J^3_1 \neq 0$ and all others
vanishing. These can be gauged away using the $U(1)$'s associated
to the D2 worldvolumes, as expected, therefore the BPS
configuration is completely determined by the frozen fields, and
the moduli space is a point.

Probing this system with a D0 brane, we find that $\partial W = 0$
implies \begin{equation*} \begin{split} z_1 a^2_3 = a^1_3 j^2_1 \\
z_2 a^1_3 = z_3 a^1_2 \\ z_1 a^3_2 = a^1_2 j^3_1. \end{split}
\end{equation*} The Kahler structure is determined by the D-terms, \begin{equation*} \begin{split}
&|J^2_1|^2 + |J^3_1|^2 - |a^1_3|^2 - |a^1_2|^2 = r_1 \\ &|J^2_1|^2
+ |a^2_3|^2 = r_2 \\ &|J^3_1|^2 + |a^3_2|^2 = r_3, \end{split}
\end{equation*} where we have absorbed the contributions of the
frozen degrees of freedom associated to noncompactness into the FI
parameters. Completing the Kahler quotient by $U(1)^3$, one
obtains the toric web diagram shown in Figure ~\ref{blowup2}.

\begin{figure}
\begin{center}
$\begin{array}{c@{\hspace{1in}}c}
 \epsfig{file=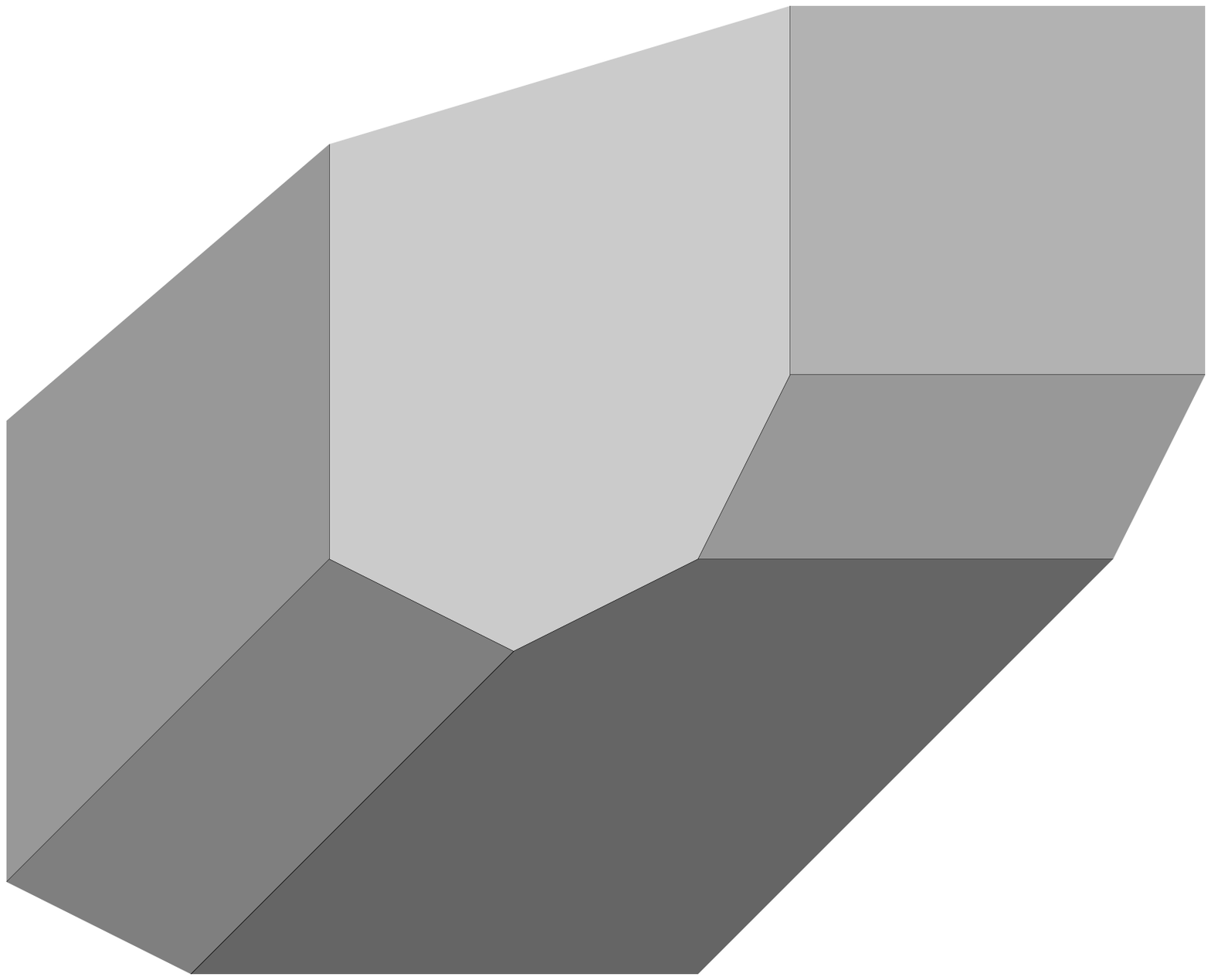,width=5cm} & \epsfig{file=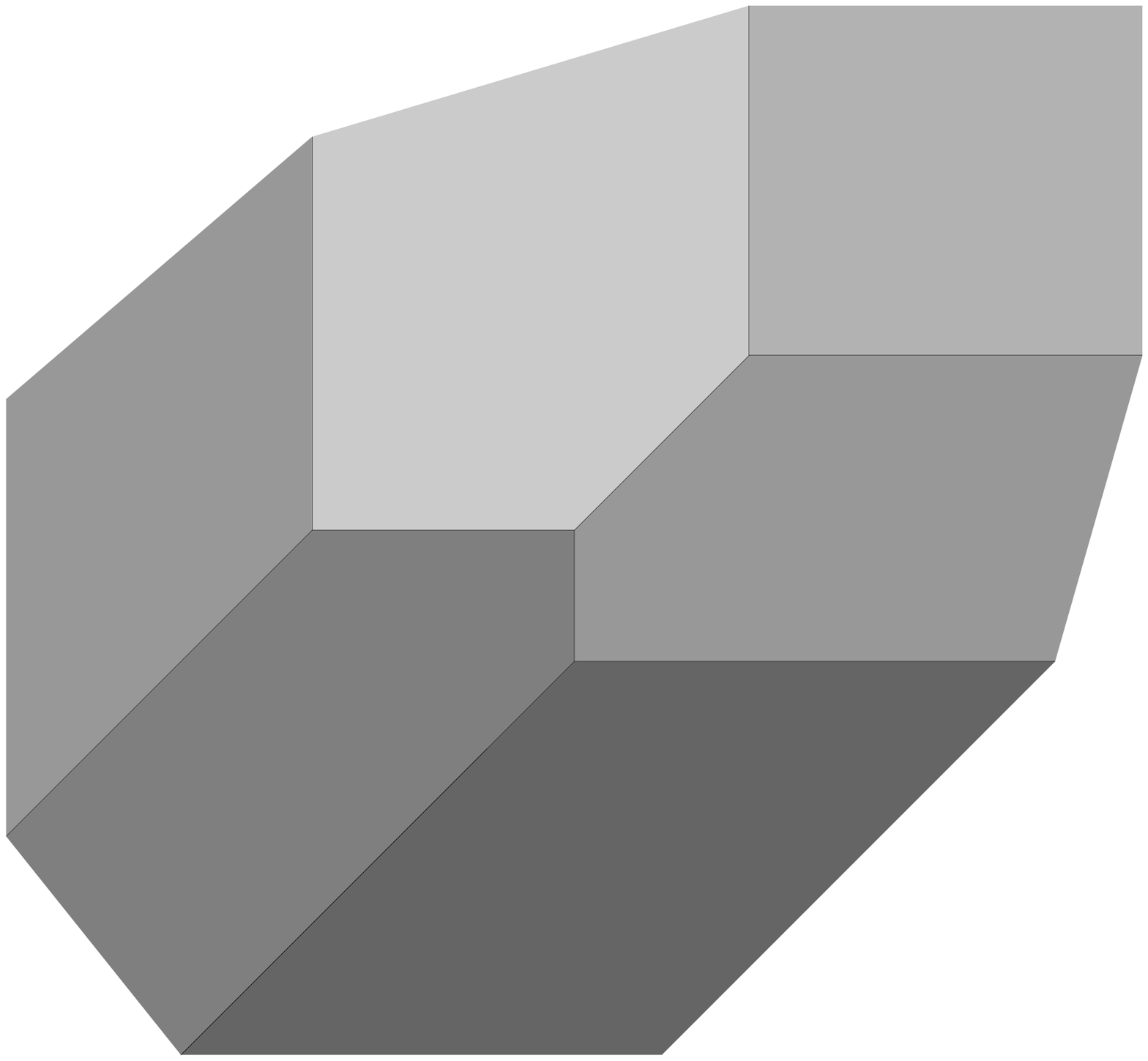,width=5cm}
  \end{array}$ \caption{Toric diagrams of some
blowups associated to $C_{\Box \Box \Box}$, related by flops.}
\label{blowup2}
\end{center}
\end{figure}

As we have seen before, the fact that the Euler character is 4
instead of the expected 3 is not a contradiction, since if $r_1 >
0$, than one of the $J^a_1 \neq 0$, and the central $\P^1$ gets
shrunken to a point. It is reasonable to expect that this is not a
wall of marginal stability, but rather a change in the attractor
tree pattern, since the $J^a_1 = 0$ branch of the moduli space for
$r_1 < 0$ is connected to the configurations in which the 2-branes
are separated. If that was the case, then a full analysis of the
fermionic terms in the matrix model would show that the anti-ghost
bundle is not the tangent bundle in this case, and only
contributes 1 from the resolved $\P^1$. This seems more natural
than a true jump in the index, because the $r_1 > 0$ classical
moduli space is singular, and one would expect an Euler character
of 2 from the conifold singularity if we were using the tangent
bundle.

The other resolutions of the blow up are also related by changing
the pattern of the attractor flow, condensing first one of the
other 2-6 tachyons. These are related geometrically via flops
through the symmetric resolution shown in Figure ~\ref{blowup2}.

\end{subsection}

\end{section}

\begin{section}{Conclusions and further directions}

We constructed matrix models possessing a topological
supersymmetry from the quiver descriptions of holomorphic branes
in a Calabi-Yau by adding the associated fermions, introducing the
multiplets resulting from motion in the Minkowski directions, and
imposing the constraints using the anti-ghost field method of
\cite{VW}. These matrix models are the topologically twisted
version of the D-brane theory in the extreme IR limit; that is,
they are theories of the BPS sector of the (unknown) dual
superconformal quantum mechanics.

The partition function was shown to localize to the Euler
character of an obstruction bundle over the classical moduli, by
proving the appropriate vanishing theorems. We evaluated the
partition function by regularizing the path integral by gauging
part of the $U(1)^3$ torus symmetry, and using the localization of
this equivariant version to the fixed points of the torus action,
giving exact agreement with known results.

The toric vertex geometry we studied can be obtained as the local
limit of various compact Calabi-Yau manifolds. The $T^3$ symmetry
guarantees that the information of the global geometry only
affects the local degrees of freedom that remain dynamical in the
limit through the values of the FI parameters and the background
VEVs of certain frozen fields. Quite generally, some gauge groups
of global quiver will be broken by the frozen VEVs, and the
associated residual off-diagonal components of the D-term can
appear in the effective matrix model.

We saw that the geometry of the BPS moduli space of a single
0-brane in a D6/D2 background in the vertex is exactly the blowup
along the curves wrapped by the 2-branes. The structure of the
moduli space of $N$ probe 0-branes was also investigated,
revealing interesting behavior. This can be viewed as a first step
in embedding the quantum foam picture of the A-model into the full
IIA theory. In particular we found that effective internal
geometry seen by 0-brane probes is indeed the blow ups of the
Calabi-Yau predicted in \cite{qfoam}.

It might be interesting to relate these ideas to the open/closed
duality discussed in \cite{Gomis}. Although the discussion there
is in the context of finding the closed string dual of Lagrangian
branes in the A-model, they find a sum over Calabi-Yau geometries
that depend on which term one one looks at in the Young diagram
expansion of the open string holonomy. From the quantum foam point
of view, these are related to the 2-brane bound states, so there
could be some connection to the effective geometries we have
investigated.

In this paper we have focused on the quiver for Donaldson-Thomas
theory in the vertex geometry, but the extension of these ideas to
many toric geometries should be straightforward. Likewise, even
the bound states of branes on a compact Calabi-Yau such as the
quintic, which have a quiver description, may be computed using
similar topological matrix models. In those cases, the technical
difficulty of finding all of the fixed points of a toric action on
the classical BPS moduli space increases proportionally. Therefore
the hope would be to find a way around performing the direct
evaluation used here.

It would be very interesting to try to apply the techniques of
matrix models at large $N$ to the quiver theories constructed
here. It is natural to expect that an intriguing structure should
emerge in the expansion about that large charge limit. This would
make sharp the idea that these matrix models give the ``CFT'' dual
of the square of the topological string, at least in the OSV
regime, since one would identify the perturbative topological
string expansion with the $1/N$ corrections in the quiver matrix
model.

Closer to the topic of the 6-brane theories studied in this paper,
one would also hope to find a large charge expansion describing
the analog of the limit shapes in \cite{topcrystal}. The quiver
theories would contain further information about the large charge
moduli space away from the torus fixed points. Such a method of
finding the asymptotic expansion of the partition function could
have implications for the entropy enigma found in \cite{DM}. On
the other hand, because our matrix models are topological, the
naive perturbative expansion is trivial, cancelling between bosons
and fermions, so novel techniques would be required to obtain the
large $N$ limit.

\end{section}

\section*{Acknowledgements}

I would like to thank Emanuel Diaconescu, Davide Gaiotto, Greg
Moore, Lubos Motl, Natalia Saulina, Alessandro Tomasiello and Xi
Yin for helpful and stimulating conversations, and especially
Robbert Dijkgraaf and Cumrun Vafa for collaboration on a related
project. I would also like to thank the Stony Brook physics
department and 4th Simons Workshop in Mathematics and Physics for
providing a stimulating environment where part of this work was
done. This research was supported in part by NSF grants
PHY-0244821 and DMS-0244464.

\end{document}